%% file: main.tex
\documentclass[a4paper,11pt]{article}
\usepackage{jheppub}

\usepackage{ifthen} % for conditional statements
\newboolean{uprightparticles}
\setboolean{uprightparticles}{false} %True for upright particle symbols

\usepackage{bm}
\makeatletter
\g@addto@macro\bfseries{\boldmath}
\makeatother

\input{lhcb-symbols-def}

\input{my-def}

\usepackage{cleveref}

\title{Mixing and \texorpdfstring{\CP}{CP} violation in \texorpdfstring{\decay{\Dz}{\RS}}{D0 -> K- pi+} decays}

\author[a,1]{Tommaso Pajero\note{Corresponding author.}}
\author[b,c]{and Michael Joseph Morello}

\affiliation[a]{Department of Physics, University of Oxford,\\Denys Wilkinson Building, Keble Road, OX1 3RH, Oxford, United Kingdom}
\affiliation[b]{Scuola Normale Superiore,\\piazza dei Cavalieri 7, 56 126, Pisa, Italy}
\affiliation[c]{INFN, sezione di Pisa,\\largo Pontecorvo 3, 56 127, Pisa, Italy}

\emailAdd{tommaso.pajero@physics.ox.ac.uk}
\emailAdd{michael.morello@sns.it}

\abstract{
We review the experimental methods to measure mixing and time-dependent \CP violation in \Dz decays into two hadrons.
While these phenomena are usually neglected for \mbox{\decay{\Dz}{\Km\pip}} decays, this approximation is not always justified.
In particular, it produces a bias on the measurement of the parameter \ycp{}, when this is performed by relying on \decay{\Dz}{\RS} decays as a normalisation channel, whose size is around 40\% of the precision of the current world average.
Finally, we estimate the sensitivity to the weak mixing phases achievable by studying \mbox{\decay{\Dz}{\RS}} and untagged \mbox{\decay{\D}{\RS}} decays, where \D stands for either of the \Dz and \Dzb mesons.
Contrary to Cabibbo-suppressed \decay{\Dz}{\hp\hm} decays, these decay channels allow to measure these phases without final-state dependent nuisance contributions from the decay amplitudes, but their sensitivity is lower by a factor of six.
}


\arxivnumber{2106.02014}

\begin{document} 
\maketitle
\flushbottom

\section{Introduction}
Charm hadrons are the only hadrons made up of solely up-type quarks where mixing and \CP violation can be studied.
Thus, they provide a unique opportunity to detect new interactions beyond the Standard Model (SM) that leave down-type quarks unaffected~\cite{Grossman:2006jg}.
Both mixing and \CP violation are highly suppressed in charm by the Glashow--Iliopoulos--Maiani mechanism and by the hierarchical structure of the Cabibbo--Kobayashi--Maskawa (CKM) matrix, and have been observed only in recent years~\cite{Aubert:2007wf,delAmoSanchez:2010xz,Lees:2012qh,Aaltonen:2013pja,Ko:2014qvu,Peng:2014oda,Staric:2015sta,LHCb-PAPER-2015-057,BaBar:2016kvp,LHCb-PAPER-2016-033,LHCb-PAPER-2017-046,LHCb-PAPER-2018-038,LHCb-PAPER-2019-001,LHCb-PAPER-2021-009,LHCb-PAPER-2021-033,LHCb-PAPER-2019-006}.
The mass and decay-width differences of the \Dz mass eigenstates are known with relative precision of around 10\%~\cite{LHCb-PAPER-2021-033}; while \CP violation in the decay amplitudes has been observed~\cite{LHCb-PAPER-2019-006}, evidence of \CP violation in \Dz mixing is still missing~\cite{Lees:2012qh,Aaltonen:2014efa,Staric:2015sta,LHCb-PAPER-2014-069,LHCb-PAPER-2016-063,LHCb-PAPER-2019-032,LHCb-PAPER-2020-045,LHCb-PAPER-2017-046,LHCb-PAPER-2019-001,LHCb-PAPER-2021-009,LHCb-PAPER-2021-033}.
The upgrades of the \lhcb experiment and the \belletwo experiment are expected to improve the precision on these observables by about a factor of 10 within the next two decades~\cite{Bediaga:2018lhg,Kou:2018nap}, allowing to attain per-cent precision on the mixing parameters, and a precision on \CP violation in the mixing comparable to the size of the SM predictions~\cite{Bigi:2011re,Bobrowski:2010xg,Kagan:2020vri,Li:2020xrz,kagan:charm2021}.

The most precise measurements of these phenomena employ multi-body decays such as \mbox{\decay{\Dz}{\KS\pip\pim}}~\cite{LHCb-PAPER-2019-001,LHCb-PAPER-2021-009}, or \Dz decays into either of the Cabibbo-suppressed (CS) final states \KK and \PP~\cite{Lees:2012qh,Staric:2015sta,LHCb-PAPER-2018-038,LHCb-PAPER-2016-063,LHCb-PAPER-2019-032,LHCb-PAPER-2020-045}, or into the wrong-sign (WS) final state \WS~\cite{LHCb-PAPER-2017-046}.\footnote{The \mbox{\decay{\Dz}{\WS}} decay is named wrong-sign since, when the \Dz flavour at production is determined by relying on the \mbox{\decay{\theDstarp}{\Dz\pip}} strong decay, the charges of the pions from the \theDstarp and \Dz decays have opposite signs.}
Measurements of two-body decays rely on right-sign (RS) \mbox{\decay{\Dz}{\RS}} decays as a normalisation channel.
Since RS decays mainly consist of Cabibbo-favoured (CF) decays without \Dz flavour oscillation, while the decay amplitude following flavour oscillation is doubly Cabibbo-suppressed (DCS), the effect of mixing on their decay rate is usually neglected in the literature.
However, it can no longer be ignored at the current level of experimental precision, and its size needs to be carefully assessed.

This is the aim of the present article, which extends the discussion in refs.~\cite{Xing:1996pn,Kagan:2020vri} by drawing the experimental consequences of the results presented therein.
The article is structured as follows.
\Cref{sect:formalism} summarises the theoretical formalism introduced in refs.~\cite{Grossman:2009mn,Kagan:2009gb,Kagan:2020vri} to parametrise mixing and time-dependent \CP violation in \Dz decays, limiting the discussion to decays into two hadrons.
The corresponding time-dependent decay rates are presented in \cref{sect:decay_rates}, where the observables $\ycp{\RS}$ and $\DY{\RS}$, analogues of the observables \ycp{f} and \DY{f} defined for the CS final states $f=\KK$ and \PP, are introduced to parametrise the first-order contributions of mixing and time-dependent \CP violation to the RS decay rate.
The parameters $\ycp{K\pi\textnormal{,untag}}$ and $\Delta Y_{K\pi}^\textnormal{untag}$ are similarly defined for untagged \decay{\D}{\RS} decays, where the flavour at production of the \Dz meson is unknown.
\Cref{sect:experiment} reviews the experimental approaches to measure the time-dependent decay rates, and pinpoints a bias equal to  $-\ycp{\RS}$ which affects most of the measurements of \ycp{f} performed to date and has been missed in the literature.
The improvement in precision that might be achieved by including the \DY{\RS} and $\Delta Y_{K\pi}^\textnormal{untag}$ observables in the experimental program of the \lhcb experiment, and by measuring \ycp{f} with untagged \D decays, are discussed as well.
\Cref{sect:conclusions} concludes by summarising the main findings.
Finally, \cref{app:quadratic} quantifies the contribution of mixing in RS decays to the parameters measured in the analysis of the time-dependent ratio of the WS to RS decays rates, and of the terms quadratic in the mixing parameters to the expansion of the time-dependent asymmetry of the decay rates of \Dz and \Dzb mesons into the final states \KK or \PP.
The former contribution might become relevant at the \lhcb upgrades.

\section{Formalism}
\label{sect:formalism}
This section reviews the theoretical formalism introduced in refs.~\cite{Grossman:2009mn,Kagan:2009gb,Kagan:2020vri} to parametrise mixing and time-dependent \CP violation in \Dz decays.

\subsection{\texorpdfstring{\Dz}{D0} mixing}
An arbitrary linear combination of \Dz and \Dzb flavour eigenstates,
\begin{equation}
    a(t)\ket{\Dz} + b(t)\ket{\Dzb},
\end{equation}
evolves in time according to the Schr\"{o}dinger equation,
\begin{equation}
    \label{eq:schrodinger}
    i\frac{\mathrm{d}}{\mathrm{d}t}
    \begin{pmatrix}
    a(t) \\ b(t)
    \end{pmatrix} =
    \bm{H}
    \begin{pmatrix}
    a(0) \\ b(0)
    \end{pmatrix},
\end{equation}
where $\bm{H}$ is the \mbox{$2 \times 2$} effective Hamiltonian governing the dynamics of the $\textnormal{\Dz--\Dzb}$ system.\footnote{Natural units are employed throughout, $\hbar=c=1$.}
This Hamiltonian can be expressed in terms of two Hermitian matrices as \mbox{$\bm{H} \equiv \bm{M} - \tfrac{i}{2}\bm{\Gamma}$}.
The \CPT symmetry, which is assumed hereafter, implies $H_{11} = H_{22} = M - \tfrac{i}{2}\Gamma$, with $M$ and $\Gamma$ the mass and decay width of the \Dz meson.
The transition amplitudes between \Dz and \Dzb mesons,
\begin{equation}
    \bra{\Dz}H\ket{\Dzb} = M_{12} - \tfrac{i}{2}\Gamma_{12}, \qquad
    \bra{\Dzb}H\ket{\Dz} = M^{\ast}_{12} - \tfrac{i}{2}\Gamma^{\ast}_{12},
\end{equation}
depend on the dispersive and absorptive mixing amplitudes, $M_{12}$ and $\Gamma_{12}$, which correspond to off-shell and on-shell transitions, respectively.
They are parametrised in terms of two \CP-even mixing parameters,
\begin{equation}
    x_{12} \equiv \frac{2 \lvert M_{12} \rvert}{\Gamma}, \qquad
    y_{12} \equiv \frac{  \lvert \Gamma_{12} \rvert}{\Gamma},
\end{equation}
and one \CP-odd weak phase,
\begin{equation}
    \phi_{12} \equiv \arg\left(\frac{M_{12}}{\Gamma_{12}}\right).
\end{equation}
The additional global phase shared by $M_{12}$ and $\Gamma_{12}$ depends on the phase conventions for quarks and mesons and is unobservable.
The parameters $x_{12}$ and $y_{12}$ are equal to the absolute value of normalised mass and decay-width differences of the eigenstates of $\bm{H}$, \mbox{$x \equiv \Delta m/\Gamma$} and \mbox{$y \equiv \Delta\Gamma/(2\Gamma)$}, up to corrections quadratic in $\sin\phi_{12}$.

\subsection{Time-dependent \texorpdfstring{\CP}{CP} violation}
\label{sect:time-dep-cvp}
The \Dz and \Dzb decay amplitudes into the \CP-conjugate final states $f$ and $\bar{f}$ are denoted as
\begin{equation}
    \begin{aligned}
    \af  \equiv& \bra{f}       \mathcal{H} \ket{\Dz},& \abf  \equiv& \bra{f}       \mathcal{H} \ket{\Dzb},\\
    \afb \equiv& \bra{\bar{f}} \mathcal{H} \ket{\Dz},& \abfb \equiv& \bra{\bar{f}} \mathcal{H} \ket{\Dzb},
    \end{aligned}
\end{equation}
where $\mathcal{H}$ is the \mbox{$\lvert \Delta C \rvert = 1$} effective Hamiltonian.

For CS decays into \CP eigenstates, \CP violation in the interference of the decay amplitudes with and without flavour oscillation is parametrised in terms of the following observables~\cite{Kagan:2020vri},
\begin{equation}
    \label{eq:lambda_CS}
    \begin{aligned}
        \lambda_f^{M} \equiv&
            \frac{M_{12}}{\lvert M_{12} \rvert} \frac{\af}{\abf}
            \equiv \etacp{f}\left\lvert\frac{\af}{\abf}\right\rvert e^{i\phi_f^M}, \\
        \lambda_f^{\Gamma} \equiv&
            \frac{\Gamma_{12}}{\lvert \Gamma_{12} \rvert} \frac{\af}{\abf}
            \equiv  \etacp{f}\left\lvert\frac{\af}{\abf}\right\rvert e^{i\phi_f^\Gamma},
    \end{aligned}
\end{equation}
where \etacp{f} equals $+1$ for \CP-even final states such as $\Kp\Km$ and $\pip\pim$, and $-1$ for \CP-odd final states such as $\KS\phiz$ and $\KS\omega$, and the weak phases $\phi_f^M$ and $\phi_f^\Gamma$ are quark- and meson-rephasing invariant and satisfy \mbox{$\phi_{12} = \phi_f^M - \phi_f^\Gamma$}.

The analogous parameters for the decays into the RS and WS final states \mbox{$f = \Km\pip$} and \mbox{$\bar{f} = \Kp\pim$} are
\begin{equation}
    \label{eq:lambda_Kpi}
    \begin{aligned}
        &\lambda_f^{M} \equiv \frac{M_{12}}{\lvert M_{12} \rvert} \frac{\af}{\abf}
                \equiv - \left\lvert\frac{\af}{\abf}\right\rvert e^{i(\phi_f^M - \Delta_f)},
        &&\lambda_{\bar{f}}^{M} \equiv \frac{M_{12}}{\lvert M_{12} \rvert} \frac{\afb}{\abfb}
                \equiv - \left\lvert\frac{\afb}{\abfb}\right\rvert e^{i(\phi_f^M + \Delta_f)}, \\
        &\lambda_f^{\Gamma} \equiv \frac{\Gamma_{12}}{\lvert \Gamma_{12} \rvert} \frac{\af}{\abf}
                \equiv - \left\lvert\frac{\af}{\abf}\right\rvert e^{i(\phi_f^\Gamma - \Delta_f)},
        &&\lambda_{\bar{f}}^{\Gamma} \equiv \frac{\Gamma_{12}}{\lvert \Gamma_{12} \rvert} \frac{\afb}{\abfb}
                \equiv - \left\lvert\frac{\afb}{\abfb}\right\rvert e^{i(\phi_f^\Gamma + \Delta_f)},
    \end{aligned}
\end{equation}
where $\Delta_f$, the strong-phase difference between the doubly Cabibbo-suppressed (DCS) and CF decay amplitudes, equals zero in the limit of $U$-spin symmetry,\footnote{Note that this convention for the strong phase differs from those adopted in refs.~\cite{HFLAV18} and \cite{LHCb-PAPER-2021-033}, respectively: \mbox{$\Delta_{\Km\pip} = - \delta_{K\pi} = - \delta^{K\pi}_{D} - \pi$}.} and the weak phases $\phi_f^M$ and $\phi_f^\Gamma$ again satisfy \mbox{$\phi_{12} = \phi_f^M - \phi_f^\Gamma$}, but in general differ from those of \cref{eq:lambda_CS}.
Finally, the minus sign in the right-hand side of the definitions accounts for the overall minus sign of the CKM matrix elements involved in the tree-level decay amplitudes, with respect to CS decays.

The chosen conventions imply that all of the phases $\phi_f^M$ and $\phi_f^\Gamma$ in \cref{eq:lambda_CS,eq:lambda_Kpi} equal the intrinsic dispersive and absorptive mixing phases $\phi_2^M$ and $\phi_2^\Gamma$, defined as the phases of $M_{12}$ and $\Gamma_{12}$ with respect to their dominant $\Delta U = 2$ contributions (hence the subscript ``2'')~\cite{Kagan:2020vri}, up to subleading corrections due to \CP violation in the decay.
These corrections depend on the final state, but are shared by the dispersive and absorptive weak phases: \mbox{$\delta\phi_f \equiv \phi^{M}_f - \phi^{M}_2 = \phi^{\Gamma}_f - \phi^{\Gamma}_2$}.
In the SM, the phases $\phi_2^M$ and $\phi_2^\Gamma$ are naively predicted to be of the order of $2\mrad$, even though enhancements up to one order of magnitude are not excluded~\cite{Bigi:2011re,Bobrowski:2010xg,Kagan:2020vri,Li:2020xrz}.
An upper bound of $5\mrad$ on the size of $\phi^\Gamma_2$ has been proposed recently~\cite{kagan:charm2021}.
The final-state dependent corrections, $\delta\phi_f$, are smaller than $10^{-6}$ for CF and DCS decays, and are suppressed with respect to $\phi_2^M$ and $\phi_2^\Gamma$ by one order in the $U$-spin breaking parameter $\epsilon \sim 0.4$ for CS decays~\cite{Kagan:2020vri}.
Since no evidence of time-dependent \CP violation has been reported to date, they are neglected in the current combinations of experimental results to maximise the achieved precision~\cite{LHCb-PAPER-2021-033,Kagan:2020vri,HFLAV18}.
Once an evidence is found, measurements of CS decays may be dropped from the combinations or replaced by the arithmetic average of the experimental observables measured in \DzKK and \DzPP decays, which allows to reduce final-state dependent contributions by a further factor of $\epsilon$~\cite{Kagan:2020vri}.

\section{Time-dependent decay rates}
\label{sect:decay_rates}
The time-dependent decay rates for \Dz and \Dzb mesons produced in their flavour eigenstates at time zero to decay into the final state $f$ at time $t$, indicated with $\Gamma(\decay{\Dz(t)}{f})$ and $\Gamma(\decay{\Dzb(t)}{f})$, respectively, can be obtained by substituting the definitions above in eq.~(50) of ref.~\cite{Kagan:2020vri}.
Since both mixing parameters are smaller than $1\%$~\cite{LHCb-PAPER-2021-033}, the rates are expanded up to quadratic order in $x_{12}$ and $y_{12}$.
This approximation is expected to hold with excellent precision also at the \lhcb and \belletwo upgrades as, owing to the exponential decay of the rates, candidates with $\Gamma t \lesssim 10$ do not contribute significantly to the experimental precision and are often discarded to ease the estimation of systematic uncertainties~\cite{LHCb-PAPER-2018-038,LHCb-PAPER-2020-045,LHCb-PAPER-2021-009}.
In the following, decay time is expressed in \Dz-lifetime units, $\tau \equiv \Gamma t$, to simplify the notation.
For the same reason, the normalisation factors from phase-space integration, which are common to \Dz and \Dzb decay rates for each final state $f$ and are additionally equal for the $\Km\pip$ and $\Kp\pim$ final states, are omitted.

\subsection{Cabibbo-suppressed decays}
For CS final states $f$ that are \CP eigenstates, the time-dependent decay rates are parametrised up to second order in the mixing parameters as
\begin{equation}
    \label{eq:decay_rates_cs}
    \begin{aligned}
    \Gamma(\decay{\Dz(\tau)}{f}) &\equiv
            e^{-\tau}\lvert\af\rvert^{2}
            \left(1 + c^+_f \tau + c^{\prime +}_f \tau^2\right),\\
    \Gamma(\decay{\Dzb(\tau)}{f}) &\equiv
            e^{-\tau}\lvert\abf\rvert^{2}
            \left(1 + c^-_f \tau + c^{\prime -}_f \tau^2\right),
    \end{aligned}
\end{equation}
where the coefficients $c^\pm_f$ and $c^{\prime \pm}_{f}$ are equal to
\begin{subequations}
    \label{eq:c_cs}
    \begin{align}
    c^\pm_f &\approx
            \etacp{f}\left[
            \mp x_{12}\sin\phi_{f}^{M} -y_{12}\cos\phi_{f}^{\Gamma}(1 \mp \Acpdec{f})\right], \label{eq:c_cs_linear}\\
    c^{\prime \pm}_{f} &\approx
            \frac{1}{2}\left[y_{12}^2 \pm x_{12}y_{12}\sin\phi_{12} \mp (x_{12}^2 + y_{12}^2)\Acpdec{f} \right]. \label{eq:c_cs_quadratic}
    \end{align}
\end{subequations}
In \cref{eq:c_cs}, the parameter
\begin{equation}
    a_{f}^{d} \equiv \frac{\lvert \af \rvert^{2} - \lvert \abf \rvert^{2}}
                          {\lvert \af \rvert^{2} + \lvert \abf \rvert^{2}}
            \approx 1 - \left\lvert\frac{\abf}{\af}\right\rvert
\end{equation}
is the \CP asymmetry in the decay, and terms proportional to \Acpdec{f} have been expanded to first order in the \CP-violation parameters \Acpdec{f}, $\sin\phi_{f}^{M}$ and $\sin\phi_{f}^{\Gamma}$.

The following \CP-even and \CP-odd combinations of $c^+_f$ and $c^-_f$ are more convenient as experimental observables,
\begin{subequations}
\begin{align}
    \ycp{f} &\equiv -\frac{c^{+}_{f} + c^{-}_{f}}{2}
                 \approx \etacp{f} y_{12}\cos\phi_{f}^{\Gamma},\label{eq:ycp_CS}\\
    \DY{f} &\equiv \frac{c^{+}_{f} - c^{-}_{f}}{2}
                 \approx \etacp{f}(- x_{12} \sin\phi^{M}_{f} + y_{12} a^{d}_{f}).\label{eq:DY_CS}
\end{align}
\end{subequations}
In the limit of \CP symmetry, the parameter \ycp{f} equals $y_{12}$ ($-y_{12}$) for \CP-even (\CP-odd) final states, and \DY{f} equals zero.

\subsection{Wrong-sign and right-sign decays}
\label{sect:ws_and_rs_decay_rates}
In this section, $f$ ($\bar{f}$) denotes the RS final state $\Km\pip$ (the WS final state $\Kp\pim$), and the notation follows ref.~\cite{pajero:2021} rather than ref.~\cite{Kagan:2020vri}.
The ratios of the DCS to CF branching fractions of \Dz and \Dzb mesons are denoted as
\begin{equation}
    R_f^+ \equiv \lvert \afb / \af \rvert^2, \qquad R_f^- \equiv \lvert \abf / \abfb\rvert^2;
\end{equation}
their average is denoted as
\begin{equation}
    R_f \equiv \frac{R_f^+ + R_f^-}{2}
\end{equation}
and is equal to \mbox{$(3.43 \pm 0.02) \times 10^{-3}$}~\cite{LHCb-PAPER-2021-033}.
Finally, the \CP asymmetry in the decay for CF decays is denoted as
\begin{equation}
    \label{eq:acp_d_CF}
    \Acpdec{f} \equiv \frac{\lvert \af \rvert^{2} - \lvert \abfb \rvert^{2}}
                          {\lvert \af \rvert^{2} + \lvert \abfb \rvert^{2}}
            \approx 1 - \left\lvert\frac{\abfb}{\af}\right\rvert,
\end{equation}
and its analogue for DCS decays as
\begin{equation}
    \label{eq:acp_d_DCS}
    \Acpdec{\bar{f}} \equiv \frac{\lvert \afb \rvert^{2} - \lvert \abf \rvert^{2}}
                          {\lvert \afb \rvert^{2} + \lvert \abf \rvert^{2}}
            \approx 1 - \left\lvert\frac{\abf}{\afb}\right\rvert.
\end{equation}
Both of these asymmetries are expected to be beyond experimental reach in the SM~\cite{Grossman:2006jg}.\footnote{As a consequence, the following equality is expected to hold, \mbox{$R^+_f = R^-_f = R_f$}. Allowing for \CP violation in the decay, instead, these parameters are related as \mbox{$R^\pm_f \approx R_f [1 \pm (\Acpdec{\bar{f}}-\Acpdec{f})]$} up to first order in \Acpdec{\bar{f}} and \Acpdec{f}. The parameter \mbox{$A_D \equiv (R^+_f - R^-_f)/(R^+_f + R^-_f)$} employed in the fits performed by the HFLAV collaboration~\cite{HFLAV18} is equal to \mbox{$\Acpdec{\bar{f}}-\Acpdec{f}$} up to first order in \Acpdec{\bar{f}} and \Acpdec{f}.}

The time-dependent decay rates of RS decays are parametrised as
\begin{equation}
    \label{eq:decay_rates_rs}
    \begin{aligned}
        \Gamma(\decay{\Dz(\tau)}{f}) &\equiv
            e^{-\tau}\lvert\af\rvert^{2}
            \left(1 + \sqrt{R_f}c^+_f \tau + c^{\prime +}_f \tau^2\right),\\
    \Gamma(\decay{\Dzb(\tau)}{\bar{f}}) &\equiv
            e^{-\tau}\lvert\abfb\rvert^{2}
            \left(1 + \sqrt{R_f}c^-_f \tau + c^{\prime -}_f \tau^2\right),
    \end{aligned}
\end{equation}
up to second order in the mixing parameters, where the coefficients $c^\pm_f$ and $c^{\prime \pm}_{f}$ are equal to
\begin{equation}
    \label{eq:c_rs}
    \begin{aligned}
    c^\pm_f &\approx
                \Big[1 \mp \tfrac{1}{2}(a^d_{\bar{f}} + a^d_f)\Big]
                (- x_{12}\cos\phi_{f}^{M}\sin\Delta_f + y_{12}\cos\phi_{f}^{\Gamma}\cos\Delta_f) \\
                &\hspace{5.5cm} \pm x_{12}\sin\phi_f^M\cos\Delta_f \pm y_{12} \sin\phi_f^\Gamma \sin\Delta_f, \\
    c^{\prime \pm}_f &\approx
            \tfrac{1}{4} (y_{12}^2 - x_{12}^2)
            + \tfrac{1}{4} R_f(1 \mp a^d_{\bar{f}} \mp a^d_f)(x_{12}^2 + y_{12}^2)
            \pm \tfrac{1}{2}R_f x_{12} y_{12}\sin\phi_{12}.
    \end{aligned}
\end{equation}
In \cref{eq:c_rs}, terms multiplying $a^d_f$ or $a^d_{\bar{f}}$ have been expanded to first order in the \CP-violation parameters $a^d_f$, $a^d_{\bar{f}}$, $\sin\phi_{f}^{M}$ and $\sin\phi_{f}^{\Gamma}$.

Employing the same approximations as above, the time-dependent decay rates of WS decays are parametrised as
\begin{equation}
    \label{eq:decay_rates_ws}
    \begin{aligned}
    \Gamma(\decay{\Dz(\tau)}{\bar{f}}) &\equiv
            e^{-\tau}\lvert\af\rvert^{2}
            \left(R_f^+ + \sqrt{R_f^+} c^{+}_{\bar{f}} \tau + c^{\prime +}_{\bar{f}} \tau^2\right),\\
    \Gamma(\decay{\Dzb(\tau)}{f}) &\equiv
            e^{-\tau}\lvert\abfb\rvert^{2}
            \left(R_f^- + \sqrt{R_f^-} c^{-}_{\bar{f}} \tau + c^{\prime -}_{\bar{f}} \tau^2\right),
    \end{aligned}
\end{equation}
where the coefficients $c^\pm_{\bar{f}}$ and $c^{\prime \pm}_{\bar{f}}$ are approximately equal to
\begin{equation}
    \label{eq:c_ws}
    \begin{aligned}
    c^{\pm}_{\bar{f}} &\approx
            (1 \mp a_f^d)(x_{12}\cos\phi_{f}^{M}\sin\Delta_f + y_{12}\cos\phi_{f}^{\Gamma}\cos\Delta_f) \\
            &\hspace{5.5cm} \pm x_{12}\sin\phi_f^M \cos\Delta_f \mp y_{12}\sin\phi_f^\Gamma\sin\Delta_f, \\
    c^{\prime \pm}_{\bar{f}} &\approx
            \tfrac{1}{4}[(x_{12}^2 + y_{12}^2)(1 \mp 2 a_f^d) \pm 2 x_{12} y_{12}\sin\phi_{12}]
            + \tfrac{1}{4}R_f^{\pm} (y_{12}^2 - x_{12}^2).
    \end{aligned}
\end{equation}

The choice of the parametrisation in \cref{eq:decay_rates_rs,eq:decay_rates_ws} ensures that the coefficients $c^\pm_{f,\bar{f}}$ and $c^{\prime\pm}_{f,\bar{f}}$ are of similar size to those in \cref{eq:decay_rates_cs}.
The suppression (enhancement) of the terms linear (and quadratic) in the mixing parameters with respect to the constant ones for RS (WS) decays, due to the different size of the CF and DCS decay amplitudes contributing to the decays without and following mixing (or vice versa), are taken into account by the $\sqrt{R_f^{(\pm)}}$ factors.
These factors equal unity up to corrections smaller than $10^{-3}$ for CS decays, where the same amplitude contributes to both the decays without and following mixing in the limit of no \CP violation.

Contrary to CS decays, where the only strong-phase difference between the amplitudes without and following mixing is given by the imaginary unit that multiplies $M_{12}^{(*)}$ in the solution of \cref{eq:schrodinger}, WS and RS decays receive a nontrivial contribution also from the strong-phase difference, $\Delta_f$, between the DCS and CF decay amplitudes.
Thus, WS and RS decays are sensitive to the weak phase $\phi^\Gamma_f$ through sine functions, unlike for CS decays, which are sensitive only to $\sin\phi^M_f$ and to $\cos\phi^\Gamma_f$.
However, this sensitivity is limited by smallness of $\Delta_f$, which is equal to $-0.17 \pm 0.07\rad$~\cite{LHCb-PAPER-2021-033}.

In analogy with \cref{eq:ycp_CS,eq:DY_CS}, the following parameters are defined for RS decays,
\begin{subequations}
\begin{align}
    \ycp{f} &\begin{aligned}[t]
            &\equiv - \sqrt{R_f} \, \frac{c^{+}_{f} + c^{-}_{f}}{2}\\
            &\approx \sqrt{R_f}\big(
                x_{12}\cos\phi_{f}^{M}\sin\Delta_f - y_{12}\cos\phi_{f}^{\Gamma}\cos\Delta_f
                \big),
            \end{aligned}\label{eq:ycp_RS}\\
    \DY{f} &\begin{aligned}[t]
            &\equiv \sqrt{R_f}\,\frac{c^{+}_{f} - c^{-}_{f}}{2} \\
            &\approx \sqrt{R_f}\big[
                 x_{12}\sin\phi_f^M\cos\Delta_f + y_{12} \sin\phi_f^\Gamma \sin\Delta_f
            + \tfrac{1}{2}(a^d_{\bar{f}} + a^d_f) (x_{12}\sin\Delta_f
            - y_{12}\cos\Delta_f) \big].
    \end{aligned}\label{eq:DY_RS}
\end{align}
\end{subequations}
In the limit of no \CP violation in the decay and of small strong phase $\Delta_f$, these parameters are equal to the negative of the parameters \ycp{f} and \DY{f} of CS decays, multiplied by a suppression factor of $\sqrt{R_f}$.

Analogous parameters are defined also for untagged decays, for which the \Dz flavour at production is unknown and the final states $f$ and $\bar{f}$ are analysed independently of the meson from which they originate,
\begin{subequations}
\begin{align}
    \ycp{K\pi\textnormal{,untag}} &\begin{aligned}[t]
            &\equiv -\frac{1}{1+R_f}
                \Big(\sqrt{R_f} \, \frac{c^{+}_{f} + c^{-}_{f}}{2}
                    + \frac{\sqrt{R^+_f}c^{+}_{\bar{f}} + \sqrt{R^-_f}c^{-}_{\bar{f}}}{2}\Big)\\
            &\approx - \frac{2 \sqrt{R_f}}{1 + R_f}\,y_{12}\cos\phi_f^\Gamma\cos\Delta_{f},
            \end{aligned}\label{eq:ycp_untag}\\
    \Delta Y_{K\pi}^\textnormal{untag} &\begin{aligned}[t]
            &\equiv \frac{  \sqrt{R_f}c^{+}_{f} + \sqrt{R^-_f}c^{-}_{\bar{f}}
                          - \sqrt{R_f}c^{-}_{f} - \sqrt{R^+_f}c^{+}_{\bar{f}}}{2 (1 + R_f)}\\
            & \approx \frac{\sqrt{R_{f}}}{1 + R_f} y_{12} [
                2 \sin\phi_{f}^\Gamma\sin\Delta_f
                + (\Acpdec{\bar{f}} - \Acpdec{f})\cos\Delta_f].
    \end{aligned}\label{eq:DY_untag}
\end{align}
\end{subequations}
In the limit of no \CP violation in the decay, $\DY{f}$ and $\DY{\kaon\pion}$ are equal to the time-integrated asymmetries of eqs. (A1) and (A2) of ref.~\cite{Kagan:2020vri}.\footnote{The same formulas had been previously presented in the phenomenological parametrisation in section V.A of ref.~\cite{Xing:1996pn}.}

\section{Experimental methods}
\label{sect:experiment}
The following sections discuss how the observables introduced so far can be measured.

\subsection{Cabibbo-suppressed decays}
\label{sect:cs_measurements}
The $c^\pm_f$ coefficients in \cref{eq:c_cs_linear} have been measured at the \B factories by modelling the CS decay rates with an exponential function~\cite{Lees:2012qh,Staric:2015sta},
\begin{equation}
    \label{eq:effective_widths}
    \begin{aligned}
        \Gamma(\decay{\Dz(\tau)}{f}) &= \lvert \af \rvert^{2} \exp(-\hat{\Gamma}_{\decay{\Dz }{f}}\tau),\\
        \Gamma(\decay{\Dzb(\tau)}{f}) &= \lvert \abf \rvert^{2} \exp(-\hat{\Gamma}_{\decay{\Dzb}{f}}\tau).
    \end{aligned}
\end{equation}
Neglecting terms quadratic in the mixing parameters, the effective decay widths defined in \cref{eq:effective_widths} satisfy
\begin{equation}
    \label{eq:effective_width}
    \hat{\Gamma}_{\decay{\Dz/\Dzb}{f}} \approx 1 - c^{\pm}_f.
\end{equation}
Under this approximation, the observables \ycp{f} and \DY{f} defined in \cref{eq:ycp_CS,eq:DY_CS} can be measured as
\begin{subequations}
\begin{align}
    \ycp{f} &\approx \frac{\hat{\Gamma}_{\decay{\Dz }{f}} + \hat{\Gamma}_{\decay{\Dzb}{f}}}{2} - 1,\label{eq:ycp_eff}\\
    \DY{f}  &\approx \frac{\hat{\Gamma}_{\decay{\Dzb}{f}} - \hat{\Gamma}_{\decay{\Dz }{f}}}{2}.\label{eq:DY_eff}
\end{align}
\end{subequations}

Precision measurements of the effective decay widths require measuring precisely both the decay time and the decay width $\Gamma$, as they enter their definition in \cref{eq:effective_widths} through~$\tau$.
Relative errors on the measurement of decay time or $\Gamma$ provoke relative errors of the same size on the measurement of the asymmetry parameter \DY{f} (unless the bias on the measurement of decay time differs for \Dz and \Dzb mesons), and thus affect it only marginally, as its value is still compatible with zero.
On the contrary, they are added to the value of \ycp{f}, which quantifies deviations of the \CP-averaged effective decay width from unity.
Therefore, they can cause large biases on this small-sized parameter.
To reduce the size of possible biases, determinations of \DY{f} and \ycp{f} based on the measurement of the effective decay widths are usually normalised to the measurement of the effective decay width of RS decays, whose deviation from unity is smaller approximately by a factor of \mbox{$\sqrt{R_{\RS}} = (5.87 \pm 0.02)\%$}~\cite{LHCb-PAPER-2021-033} with respect to that of CS decays:
\begin{equation}
    \hat{\Gamma}_{\decay{\Dz}{\RS}/\decay{\Dzb}{\WS}} \approx 1 - \sqrt{R_{\RS}}c^{\pm}_{\RS}.
\end{equation}
The \cref{eq:ycp_eff,eq:DY_eff} are modified accordingly, yielding
\begin{subequations}
\begin{align}
    \ycp{f} - \ycp{\RS} &\approx
            \frac{\hat{\Gamma}_{\decay{\Dz }{f  }} + \hat{\Gamma}_{\decay{\Dzb}{f  }}}
                 {\hat{\Gamma}_{\decay{\Dz }{\RS}} + \hat{\Gamma}_{\decay{\Dzb}{\WS}}} - 1,\label{eq:ycp_eff_wrt_RS}\\
    \frac{\DY{f}}{1 + \ycp{\RS}} &\approx
            \frac{\hat{\Gamma}_{\decay{\Dzb}{f  }} - \hat{\Gamma}_{\decay{\Dz }{f  }}}
                 {\hat{\Gamma}_{\decay{\Dz }{\RS}} + \hat{\Gamma}_{\decay{\Dzb}{\WS}}},\label{eq:DY_eff_wrt_RS}
\end{align}
\end{subequations}
where the dependence on $\Gamma$ and on the time scale cancels out in the ratio of the effective decay widths.
A similar cancellation has been exploited by the \lhcb and \belle collaborations, which measured the asymmetry of the effective lifetimes of \Dz and \Dzb mesons~\cite{LHCb-PAPER-2013-054,Staric:2015sta},
\begin{equation}
    \label{eq:agamma}
    \agamma{f} \equiv \frac{\hat{\Gamma}_{\decay{\Dz}{f}} - \hat{\Gamma}_{\decay{\Dzb}{f}}}
                           {\hat{\Gamma}_{\decay{\Dz}{f}} + \hat{\Gamma}_{\decay{\Dzb}{f}}}
    \approx - \frac{\DY{f}}{1 + \ycp{f}}.
\end{equation}
This observable differs from that of \cref{eq:DY_eff_wrt_RS} for the different calibration channel at denominator as well as for an overall minus sign.

Given the smallness of $\ycp{f}$~\cite{HFLAV18}, the observable \agamma{f} is equal to the negative of \DY{f} within 1\% precision.
The observable defined in \cref{eq:DY_eff_wrt_RS} is equal to \DY{f} with even better precision, as \ycp{\RS} is smaller than \ycp{f} approximately by a factor of $\sqrt{R_{\RS}}$.
On the contrary, the approximation that $\hat{\Gamma}_{\decay{\Dz}{\RS}/\decay{\Dzb}{\WS}}$ is equal to unity cannot be used effectively in the measurement of \ycp{f} through \cref{eq:ycp_eff_wrt_RS}, since the contribution from \ycp{\RS} in the left-hand side yields relative corrections to \ycp{f} of the order of $\sqrt{R_{\RS}}\approx 6\%$~\cite{LHCb-PAPER-2021-033}.

In recent years, alternative methods have been preferred to measure \DY{f} and \ycp{f}.
Measuring the effective decay widths from the decay time distributions requires knowing with high precision the reconstruction efficiency as a function of decay time, even when a normalisation channel such as \DzRS is available.
This task is challenging in case of stringent trigger requirements on \Dz displacement-related variables, such as those set at hadron colliders.
However, it can be mitigated by measuring the parameter \DY{f} from the slope of the time-dependent asymmetry of the decay rates of \Dz and \Dzb mesons into the final state $f$~\cite{Aaltonen:2014efa,LHCb-PAPER-2014-069,LHCb-PAPER-2016-063,LHCb-PAPER-2019-032,LHCb-PAPER-2020-045},
\begin{equation}
    \label{eq:acp_CS}
    A_{\CP}(f,\tau) \equiv \frac{\Gamma(\decay{\Dz(\tau)}{f}) - \Gamma(\decay{\Dzb(\tau)}{f})}
                             {\Gamma(\decay{\Dz(\tau)}{f}) + \Gamma(\decay{\Dzb(\tau)}{f})}\\
                \approx \Acpdec{f} + \DY{f}\tau,
\end{equation}
where terms quadratic in the mixing parameters or in $\Acpdec{f}$ have been neglected.
The average reconstruction efficiency of \Dz and \Dzb decays cancels out in the ratio, and only charge-dependent efficiency effects must be corrected for.
In addition, the \CP-even terms quadratic in the mixing parameters in \cref{eq:c_cs_quadratic} cancel out in the numerator and do not bias the results, while the \CP-odd quadratic terms
are negligible within the current precision as well as that attainable at the \lhcb upgrades and at the \belletwo experiment, as shown in \cref{app:quadratic}.
The measurements based on the effective decay widths do not benefit from this feature: the \CP-even terms of the $c^{\prime \pm}_f$ coefficients in \cref{eq:c_cs_quadratic} can be of the same order of the \CP-odd terms of the $c^\pm_f$ coefficients in \cref{eq:c_cs_linear}, thus limiting the validity of the assumption in \cref{eq:effective_width}.

The \ycp{f} parameter has analogously been measured from the time-dependent ratio of the decay rates of CS to RS decays~\cite{LHCb-PAPER-2018-038},
\begin{equation}
    \label{eq:ycp_from_ratio}
    \begin{aligned}
    &\frac{\Gamma(\decay{\Dz(\tau)}{f}) + \Gamma(\decay{\Dzb(\tau)}{f})}
         {\Gamma(\decay{\Dz(\tau)}{\Km\pip}) + \Gamma(\decay{\Dzb(\tau)}{\Kp\pim})}
         \approx \textnormal{const.} \times \Big\{1 - (\ycp{f} - \ycp{\RS})\tau \\
    &\hspace{4.5cm} + \big[\tfrac{1}{4}(x_{12}^2 + y_{12}^2)(1 - R_{\RS})
    + \ycp{\RS}(\ycp{\RS} - \ycp{f})\big]\tau^2\Big\},
    \end{aligned}
\end{equation}
where terms cubic in the mixing parameters have been neglected and terms proportional to \Acpdec{f} and \Acpdec{\bar{f}} have been expanded to first order in the \CP-violation parameters \Acpdec{f}, \Acpdec{\bar{f}}, $\sin\phi^M_f$ and $\sin\phi^\Gamma_f$.
The negative of the slope differs from \ycp{f} by the same quantity, $-\ycp{\RS}$, as the ratio of the effective decay widths in \cref{eq:ycp_eff_wrt_RS}, and cannot be neglected.
The term quadratic in the mixing parameters is negligible at the current level of experimental precision, but will become relevant at large values of $\tau$ in the expected measurements employing the \lhcb Run~ 2 data sample (2015--2018) and beyond.
When this happens, approximating the time evolution of the ratio with an exponential, as done in ref.~\cite{LHCb-PAPER-2018-038}, might provoke biases to the measurement, as the quadratic term is independent of the linear one, though accidentally its numerical value is not so different from that expected from the expansion of the exponential, $\tfrac{1}{2}(\ycp{f} - \ycp{\RS})^2$~\cite{LHCb-PAPER-2021-033}.

\begin{figure}[tb]
    \begin{center}
    \includegraphics[width=0.485\linewidth]{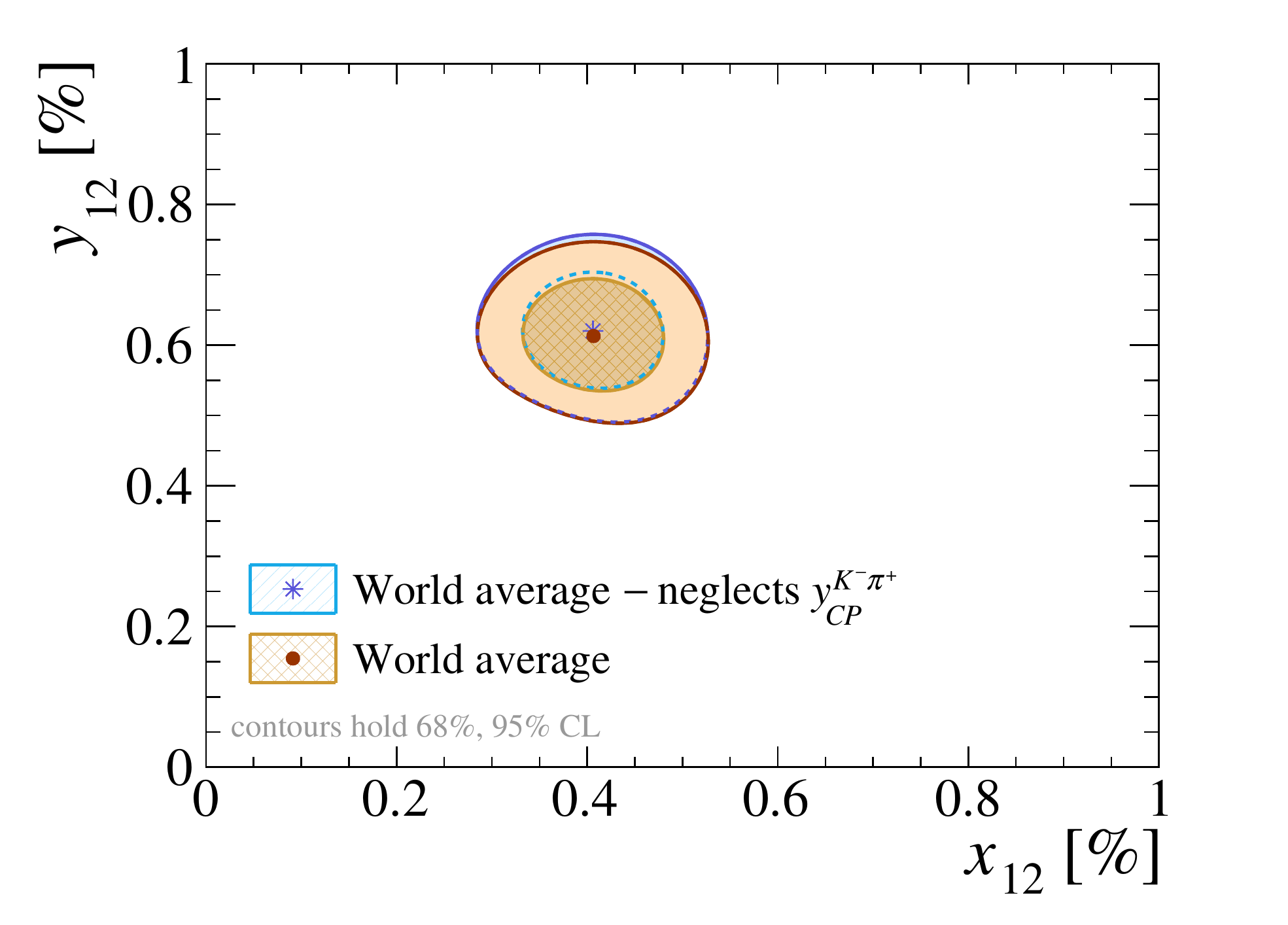}
    \includegraphics[width=0.485\linewidth]{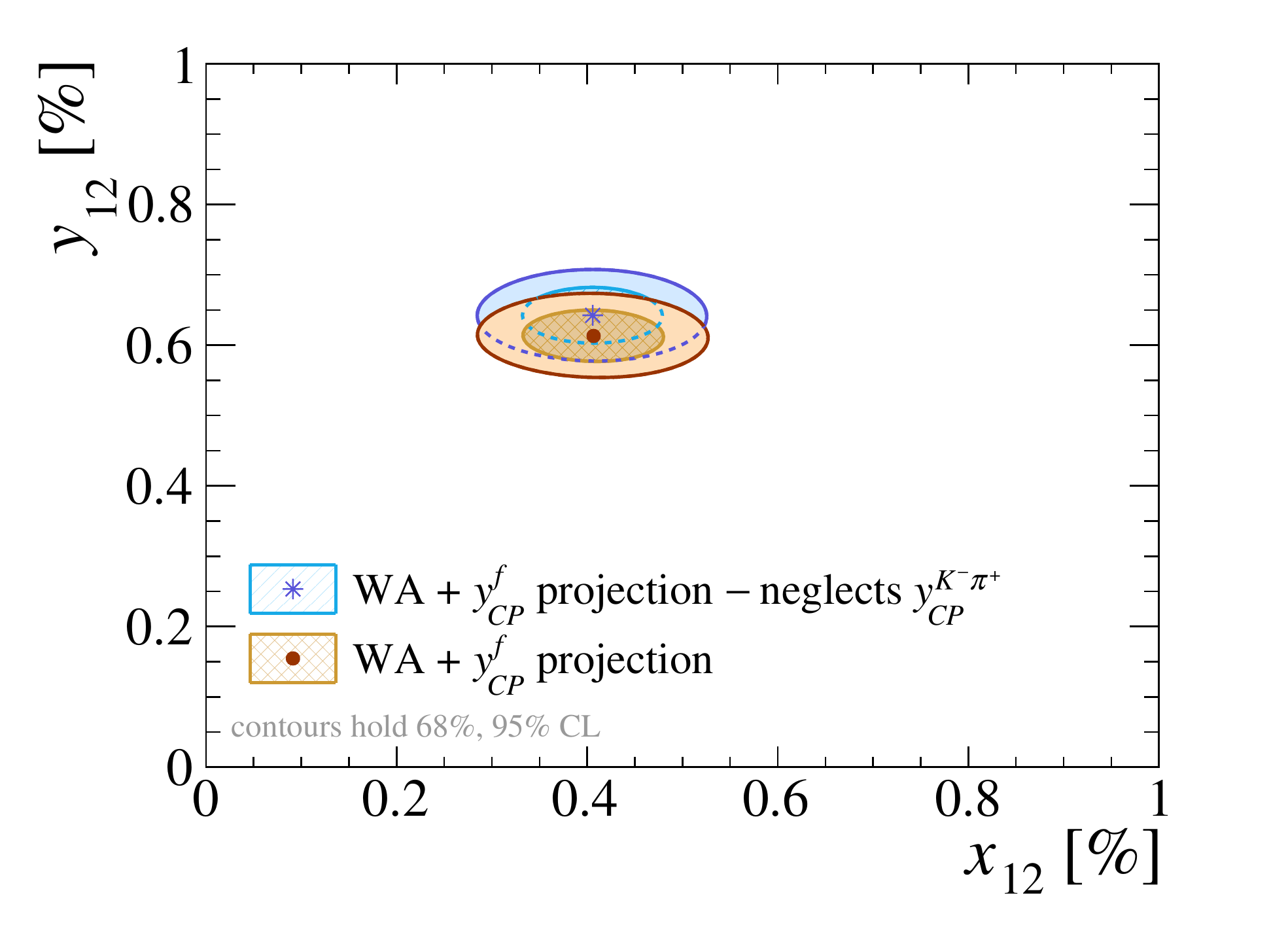}
    \vspace*{-0.7cm}
    \end{center}
    \caption{
        Impact of neglecting the \ycp{\RS} terms in \cref{eq:ycp_eff_wrt_RS,eq:ycp_from_ratio} on the world average of the mixing parameters, (left) as of today and (right) after including a new measurement of \ycp{f}, where $f$ equals \KK or \PP, with the precision achievable with the \theDstarp-tagged data sample collected by the LHCb experiment during its Run~2 data taking (2015--2018).
        In the right plot, the value of \ycp{f} is fixed to its point estimate based on the results of the fit in ref.~\cite{HFLAV18}, and its uncertainty is set to $0.3 \times 10^{-4}$.
        Measurements of the $\gamma$ angle of the CKM unitary triangle relying on two-body \Dz decays, which improve the precision on the strong-phase $\Delta_{\RS}$ and indirectly on the current world average of $y_{12}$ by around 40\%~\cite{LHCb-PAPER-2021-033}, are not included for simplicity.
        The code employed to perform the fits is available on GitHub~\cite{pajero:code}.
    }
    \label{fig:ycp-bias}
\end{figure}

The term \ycp{\RS} in the left-hand side of \cref{eq:ycp_eff_wrt_RS} and in the linear term on the right-hand side of \cref{eq:ycp_from_ratio} has been neglected in all combinations of charm mixing and \CP-violation measurements~\cite{HFLAV18,Kagan:2020vri} except for that in ref.~\cite{LHCb-PAPER-2021-033}, but amounts to about 40\% of the precision of the current world average of \mbox{$\ycp{} = (7.19 \pm 1.13)\times 10^{-3}$}~\cite{HFLAV18}, where the convention $\etacp{f} = 1$ is employed and measurements of \CP-odd final states have been multiplied by $-1$.
Since the average is driven by measurements of \CP-even final states, it is biased towards larger values.
While its current value contributes marginally to the results of the combinations, the statistical uncertainty attainable with the data collected during the Run 2 of the \lhcb experiment is smaller by a factor of 5.
When this precision is achieved, correcting the expressions used in the fits according to \cref{eq:ycp_eff_wrt_RS,eq:ycp_from_ratio} will be indispensable to avoid large biases, as shown in \cref{fig:ycp-bias}.

\subsection{Wrong-sign decays}
\label{sect:ws_measurements}
In the measurements of WS decays into the final state $\bar{f}=\WS$, modelling the exponential factor $e^{-\tau}$ in the decay rates of \cref{eq:decay_rates_ws}, as well as the time dependence of the efficiency due to trigger and reconstruction effects, the \D-meson production asymmetry and most of the detection asymmetries, is usually avoided by analysing the ratio of the WS to RS decay rates~\cite{Aubert:2007wf,Ko:2014qvu,Aaltonen:2013pja,LHCb-PAPER-2016-033,LHCb-PAPER-2017-046},
\begin{equation}
    \label{eq:WS_to_RS}
    \begin{aligned}
    \frac{\Gamma(\decay{\Dz(\tau)}{\bar{f}})}{\Gamma(\decay{\Dz(\tau)}{f})}
        &\approx R_f^+ + \sqrt{R_f^+} c^{+}_{\bar{f}} \tau + c^{\prime +}_{\bar{f}} \tau^2, \\
    \frac{\Gamma(\decay{\Dzb(\tau)}{f})}{\Gamma(\decay{\Dzb(\tau)}{\bar{f}})}
        &\approx R_f^- + \sqrt{R_f^-} c^{-}_{\bar{f}} \tau + c^{\prime -}_{\bar{f}} \tau^2.
    \end{aligned}
\end{equation}
Here, the contribution of RS decays to the linear and quadratic terms is suppressed by a relative factor of $R_f$ and is neglected.
It might become relevant at the future upgrades of the \lhcb experiment, but only for the \CP-even observables, as detailed in \cref{app:quadratic}.

An alternative approach consists in taking the ratio of the WS decay rates for a given \Dz initial flavour, over RS decay rates from \Dz mesons with opposite initial flavour. This new observable, $\Gamma(\decay{\Dz(\tau)}{\bar{f}}) / \Gamma(\decay{\Dzb(\tau)}{\bar{f}})$, is indistinguishable from that in \cref{eq:WS_to_RS} up to a multiplicative factor of $(1\pm\Acpdec{f})/(1\mp\Acpdec{f}) \approx 1 \pm 2\Acpdec{f}$, where the top (bottom) sign applies to WS decays of \Dz (\Dzb) mesons, but presents different experimental advantages and challenges.
The approximate cancellation in the ratio of the production asymmetry and of the detection asymmetry of the accompanying particles employed to infer the \Dz flavour at production (typically a \pip meson from a \decay{\theDstarp}{\Dz\pip} decay or a muon from a \decay{\Bb}{\Dz\mun X} decay, where $X$ stands for an arbitrary number of unreconstructed particles) does not take place any longer.
On the other hand, the detection asymmetry of the $\Kpm\pimp$ final state approximately cancels out in the ratio, whereas it needs to be corrected for when using the standard observable of \cref{eq:WS_to_RS}.
Depending on the strategy adopted to cancel the asymmetries, one or other of the two observables could be more convenient.

\subsection{Right-sign decays}
\label{sect:rs_measurements}
All aforementioned measurements employ RS \decay{\Dz}{f} decays, where $f$ stands for $\RS$, merely as a normalisation channel, whose decay width is used as a proxy for $\Gamma$.
However, RS decays are sensitive to \CP violation in the mixing, although with lower sensitivity with respect to CS and WS decays.
A convenient observable to measure this effect is the time-dependent asymmetry between the \Dz and \Dzb RS decay rates,
\begin{equation}
    \label{eq:acp_RS}
    A_{\CP}(f,\tau) \equiv \frac{\Gamma(\decay{\Dz(\tau)}{f}) - \Gamma(\decay{\Dzb(\tau)}{\bar{f}})}
                             {\Gamma(\decay{\Dz(\tau)}{f}) + \Gamma(\decay{\Dzb(\tau)}{\bar{f}})}
    \approx a_{f}^{d} + \Delta Y_f\tau,
\end{equation}
where terms quadratic in the mixing parameters and in \Acpdec{f} have been neglected.

The relative experimental sensitivity to \CP violation in the mixing of measurements of \DY{\RS} and \DY{\textnormal{CS}} is estimated as follows.
The sensitivity of \DY{\RS} to this phenomenon is suppressed roughly by a factor of $\sqrt{R_{\RS}}$ with respect to \DY{\textnormal{CS}}; see \cref{eq:DY_CS,eq:DY_RS}.
On the other hand, the branching fraction of \decay{\Dz}{\RS} decays is larger by a factor of about 9.7 (27.1) than that of \DzKK (\DzPP) decays~\cite{PDG2020}.
Since the precision of the measured asymmetries is proportional to $N^{-1/2}$, where $N$ is the number of reconstructed decays, the statistical uncertainty achievable for \DY{\RS} is around 37\% of that of the weighted average of \DY{\KK} and \DY{\PP}, where the reconstruction efficiency and background contamination are assumed to be the same for all three decay channels.
This estimate is confirmed by the latest measurements of \DY{\KK/\PP}, where the compatibility of the measurement of \DY{\RS} with zero is used to cross-check the effectiveness of the analysis procedure in removing nuisance asymmetries~\cite{Aaltonen:2014efa,LHCb-PAPER-2014-069,LHCb-PAPER-2016-063,LHCb-PAPER-2019-032,LHCb-PAPER-2020-045}.
The relative sensitivity to \CP violation in the mixing achievable with \DzRS with respect to \mbox{\decay{\Dz}{\KK/\PP}} decays is thus \mbox{$\sqrt{R_{\RS}} / 37\% \approx 16\%$}.
While currently it is not competitive with that of CS decays, it might become of interest once a nonzero value of \DY{\KK/\PP} is measured, since it provides cleaner access to the mixing phase $\phi^M_2$~\cite{Kagan:2020vri}.

The measurement of \DY{\RS} is effectively uncorrelated with that of the WS-to-RS ratio, in which \CP-violating contributions from RS decays are expected to be negligible at all foreseen upgrades of the \lhcb and \belletwo experiments; see \cref{app:quadratic}.
Therefore, measuring the WS-to-RS ratio and the \DY{\RS} parameter is equivalent to measuring the time-integrated asymmetries of WS and RS decays, as proposed in section V.A of ref.~\cite{Xing:1996pn} and in appendix A of ref.~\cite{Kagan:2020vri}.
However, it is more convenient experimentally, since the former observables are by construction less sensitive to time-independent nuisance asymmetries.
Subleading nuisance asymmetries can be corrected for by relying on control channels for the WS-to-RS ratio~\cite{LHCb-PAPER-2016-033,LHCb-PAPER-2017-046}, and by building on the methods developed in refs.~\cite{LHCb-PAPER-2016-063,LHCb-CONF-2019-001,LHCb-PAPER-2020-045} for the \DY{\RS} observable.
Neither of these approaches degrades significantly the statistical uncertainty of the measurement, contrary to the more complex corrections needed to measure time-integrated asymmetries; see for example ref.~\cite{LHCb-PAPER-2014-013}.

A direct consequence of the considerations above is that the compatibility of the measurement of \DY{\RS} with zero will not be utilisable any longer to check the effectiveness of the analysis method in the measurements of \DY{\KK} and \DY{\PP}, once these are measured to differ significantly from zero.
This problem might be bypassed by calculating the confidence interval for \DY{\RS} based on the previous measurements of \DY{\KK}, \DY{\PP} and other available charm measurements, which allow to determine the \CP-violation parameters $\phi^M_2$ and $\phi^\Gamma_2$ with better precision.\footnote{This relies on the assumption that the shifts of $\phi^M_f$ and $\phi^\Gamma_f$ with respect to $\phi^M_2$ and $\phi^\Gamma_2$ are small. While this condition is not necessarily verified for the \KK and \PP final states, the size of the final-state dependent shifts can be reduced by utilising the arithmetic average of \DY{\KK} and \DY{\PP}; see ref.~\cite{Kagan:2020vri}.}
Issues may arise if the size of the analysed sample increases considerably from one measurement to the subsequent, such that the precision achieved with the new sample of RS decays is comparable to that of the measurements performed with the previous data samples of CS and WS decays.
This would require an increase of the data sample size by a factor of approximately $(16\%)^{-2} \approx 40$, where 16\% is the relative sensitivity of RS and CS measurements at equal number of produced \Dz mesons, as estimated above.

\subsection{Untagged decays}
\label{sect:untagged}
The \CP asymmetry of untagged \D decays into the final state \mbox{$f=\Km\pip$} is defined, in analogy with \cref{eq:acp_RS}, as
\begin{equation}
    \label{eq:acp_Kpi_untagged}
    \begin{aligned}
    A_{\CP}^{\textnormal{untag}}(f,\tau) &\equiv \frac{[\Gamma(\decay{\Dz(\tau)}{f}) + \Gamma(\decay{\Dzb(\tau)}{f})] - 
                              [\Gamma(\decay{\Dzb(\tau)}{\bar{f}}) + \Gamma(\decay{\Dz(\tau)}{\bar{f}})]}
                             {[\Gamma(\decay{\Dz(\tau)}{f}) + \Gamma(\decay{\Dzb(\tau)}{f})] + 
                              [\Gamma(\decay{\Dzb(\tau)}{\bar{f}}) + \Gamma(\decay{\Dz(\tau)}{\bar{f}})]} \\
    &\approx a_{K\pi}^{d\textnormal{,untag}} + \Delta Y_{K\pi}^\textnormal{untag}\,\tau,
    \end{aligned}
\end{equation}
where the \CP asymmetry in the decay is equal to
\begin{equation}
    a_{K\pi}^{d\textnormal{,untag}} \approx \Acpdec{f} - R_{f} \Acpdec{\bar{f}},
\end{equation}
$\Delta Y_{K\pi}^\textnormal{untag}$ has been defined in \cref{eq:DY_untag}, and terms quadratic in the mixing parameters or in the \CP asymmetries \Acpdec{f} and \Acpdec{\bar{f}} have been neglected.

These parameters can be measured with better precision than their analogues for RS decays.
The \lhcb experiment currently provides the largest yields of \Dz decays into two charged hadrons, with the largest yields coming from mesons produced promptly in proton--proton collisions.
Since the production cross-section of \Dz mesons is larger than that of \theDstarp mesons by a factor of 2.6~\cite{LHCb-PAPER-2012-041,LHCb-PAPER-2015-041}, the branching fraction of the \theDstarp meson into the $\Dz\pip$ final state is 68\%~\cite{PDG2020}, and the reconstruction efficiency of the low-momentum pion from the \theDstarp decay is relatively low,
the ratio of the yields of untagged and tagged decays is close to 5~\cite{LHCb-CONF-2016-005}.

Measuring $a_{K\pi}^{d\textnormal{,untag}}$ is very challenging due to the need to determine the \Dz production asymmetry and to correct for the detection asymmetry of the \RS final state.
On the other hand, the slope $\Delta Y_{K\pi}^\textnormal{untag}$ might be measured analogously to \DY{\RS}, by building on the methods developed in refs.~\cite{LHCb-PAPER-2016-063,LHCb-CONF-2019-001,LHCb-PAPER-2020-045} to correct for the time-dependent component of the nuisance asymmetries.
Since the tagged sample is a subset of the untagged sample, the measurement of $\Delta Y_{K\pi}^\textnormal{untag}$ would be correlated with those of the WS-to-RS ratio and of \DY{\RS}.
However, the correlation coefficient can be measured and is expected to be small due to the very different size of the two samples.
While the precision achievable for $\Delta Y_{K\pi}^\textnormal{untag}$ is better by a factor of \mbox{$\sqrt{5} \approx 2.2$} with respect to that of \DY{\RS}, where 5 is the yields ratio estimated above, the sensitivity of $\Delta Y_{K\pi}^\textnormal{untag}$ to $\phi^\Gamma_f$ is suppressed by a factor of \mbox{$2 (y_{12}/x_{12})\tan\Delta_f \approx 0.54 \pm 0.26$}~\cite{LHCb-PAPER-2021-033} with respect to the sensitivity of \DY{\RS} to $\phi^M_f$; see \cref{eq:DY_RS,eq:DY_untag}.
The larger combinatorial background of untagged decays, which is largely rejected in tagged decays thanks to the low $Q$ value of the \theDstarp decay, might decrease the relative precision between the two samples, but all in all RS and untagged decays might provide similar sensitivity to the two weak mixing phases.

Finally, untagged \decay{\D}{\Kp\Km/\pip\pim} and \decay{\D}{K\pi} decays could be used also to measure $\ycp{} - \ycp{K\pi\textnormal{,untag}}$ with the same methods detailed in \cref{sect:cs_measurements}.
Again, this would allow to collect larger yields and to avoid the reconstruction effects of the low-momentum pion from the \theDstarp decay, at the cost of increased combinatorial background.
The shift of the measured observable from \ycp{} due to the normalisation \decay{D}{K\pi} channel would be around twice as large as that in \cref{eq:ycp_from_ratio}; see  \cref{eq:ycp_RS,eq:ycp_untag}.

\section{Summary}
\label{sect:conclusions}

The theoretical formalism to describe mixing and \CP violation in \Dz decays into two charged hadrons (\KK, \PP, \WS or \RS) has been reviewed, along with the experimental methods to measure them.
An analogous discussion employing the phenomenological parametrisation can be found in appendices A and B of ref.~\cite{pajero:2021}.
The impact of these phenomena on the RS decay rates, which has usually been neglected in the literature, has been quantified in \cref{sect:ws_and_rs_decay_rates,sect:rs_measurements}.\footnote{References~\cite{Xing:1996pn} and \cite{Kagan:2020vri} have discussed the effect of \CP violation in section V.A and in appendix A, respectively. Here, their discussion has been extended by including possible contributions from \CP violation in the decay in addition to those from \CP violation in the mixing and in the interference between mixing and decay, and by employing a different parametrisation for RS decays.}

Mixing in RS decays has been shown to affect the measurements of the parameter \ycp{} that rely on RS decays as normalisation channel, shifting the results by $-\ycp{\RS}$, where the latter observable is defined in \cref{eq:ycp_RS}; see \cref{eq:ycp_eff_wrt_RS,eq:ycp_from_ratio}.
The size of the shift is about 40\% of the uncertainty of the current world average.
This effect has been missed in the literature and is not accounted for in the combinations of the time-dependent measurements of charm decays in refs.~\cite{HFLAV18,Kagan:2020vri}.
Its inclusion in the next iterations of the combinations, as first done in ref.~\cite{LHCb-PAPER-2021-033}, will be crucial to avoid inconsistencies and biases in the fits results.

The \ycp{} parameter could be measured also by relying on untagged decays as in ref.~\cite{E791:1999bzz}, achieving larger yields at the cost of slightly increased background.
The deviation of the observable, $\ycp{} - \ycp{K\pi\textnormal{,untag}}$, from \ycp{} would be larger than that of RS decays by around a factor of two; see \cref{eq:ycp_RS,eq:ycp_untag}.

The sensitivity of RS and untagged \mbox{\decay{\D}{\RS}} decays to the weak mixing phases has been estimated in \cref{sect:rs_measurements,sect:untagged}.
While it is worse than that achievable by analysing the \KK and \PP final states by approximately a factor of 6, it might become interesting once time-dependent \CP violation is observed.
In fact, final-state dependent corrections to the weak mixing phases $\phi^M_2$ and $\phi^\Gamma_2$ are negligible for RS and WS decays, contrary to what happens for CS decays.
The two decay channels would be complementary, as RS and \decay{\D}{\Km\pip} decays mostly provide sensitivity to $\phi^M_2$ and $\phi^\Gamma_2$, respectively, with similar experimental precision, and are nearly statistically independent.

Attention should be paid to the usage of \RS decays as a control channel in the measurements of the \DY{f} parameter for CS final states $f$, as the compatibility of \DY{\RS} with zero will no longer be guaranteed once \DY{f} is measured to differ significantly from zero, or if the size of the collected sample of \Dz decays into two charged hadrons increases significantly with respect to that employed in previous measurements, as shown in \cref{sect:rs_measurements}.

\acknowledgments

We are indebted to Alexander Kagan and Luca Silvestrini, whose recent work~\cite{Kagan:2020vri} inspired this article.
We are grateful to Roberto Ribatti for discussions about the new observable proposed in \cref{sect:ws_measurements}.
The work of T.P. was supported by the Science and Technology Facility Council [grant number ST/S000933/1].

\appendix

\section{Subleading corrections}
\label{app:quadratic}
This appendix reports subleading corrections to \cref{eq:acp_CS,eq:WS_to_RS}, which might become relevant at the precision foreseen at the end of the \lhcb Upgrade~II data taking~\cite{Bediaga:2018lhg}.

The contribution of mixing and \CP violation in RS decays has been neglected in \cref{eq:WS_to_RS} for the WS-to-RS ratios.
The complete expressions up to quadratic order in the mixing parameters and to linear order in the \CP asymmetries \Acpdec{f} and \Acpdec{\bar{f}} are
\begin{equation}
    \label{eq:ws_rs_ratio_expansion}
    \begin{aligned}
    &\frac{\Gamma(\decay{\Dz(\tau)}{\bar{f}})}
          {\Gamma(\decay{\Dz(\tau)}{f      })}
    \Bigg[\frac{\Gamma(\decay{\Dzb(\tau)}{f})}
               {\Gamma(\decay{\Dzb(\tau)}{\bar{f}})}\Bigg]
        \approx R_f\Big[1 \pm (\Acpdec{\bar{f}} - \Acpdec{f})\Big] \\
        &\hspace{0.3cm} + \sqrt{R_f}\bigg[
                  x_{12} \cos\phi_f^M \sin\Delta_f \Big(1 + R_f \pm \frac{\Acpdec{\bar{f}} - 3\Acpdec{f}}{2}\Big) 
            + y_{12} \cos\phi_f^\Gamma \cos\Delta_f \Big(1 - R_f \pm \frac{\Acpdec{\bar{f}} - 3\Acpdec{f}}{2}\Big)\\
            &\hspace{8.3cm}
                \pm x_{12} \sin\phi^M_f \cos\Delta_f
                \mp y_{12} \sin\phi^\Gamma_f \sin\Delta_f
            \bigg] \tau\\
        &\hspace{0.3cm} + \bigg[
            \frac{1}{4}(x_{12}^2 + y_{12}^2)(1 \mp 2 \Acpdec{f})
            \pm \frac{1}{2}x_{12}y_{12}(\sin\phi_f^M - \sin\phi^\Gamma_f)
        \bigg] \tau^2,
    \end{aligned}
\end{equation}
where the following terms are additionally neglected: terms proportional to the \CP asymmetries \Acpdec{f} and \Acpdec{\bar{f}} and to either of the parameters $\sin\phi^M_f$, $\sin\phi^\Gamma_f$ or $R_f$; terms proportional to $\sin\phi^M_f$ or $\sin\phi^\Gamma_f$ and to $R_f$; and terms proportional to the second power of the mixing parameters and to $R_f$.
This approximation will have a negligible impact on the theoretical interpretation of the observables even at the end of the \lhcb Upgrade~II, since its target relative precision on the terms linear and quadratic in decay time is about 2\% and 0.5\%, respectively~\cite{Bediaga:2018lhg}.\footnote{This estimate assumes that the correlation between the measured parameters will be the same as in past measurements~\cite{LHCb-PAPER-2017-046}.}
Terms of third order in the mixing parameters will be equally negligible, as they are proportional to $\sqrt{R_f}$.
The contribution of RS decays to the time dependence of the ratios corresponds to the terms proportional to $R_f$ in the second line of \cref{eq:ws_rs_ratio_expansion}, and if neglected will cause a bias on the linear coefficient of the time-dependent ratio of the same order of the statistical precision targeted by the \lhcb Upgrade~II.

The analogue of \cref{eq:ws_rs_ratio_expansion} for the new experimental observable proposed in \cref{sect:ws_measurements} is just the same up to a multiplicative factor of $1 + 2\Acpdec{f}$ (of $1 - 2\Acpdec{f}$) for \mbox{$\Gamma(\decay{\Dz(\tau)}{\bar{f}}) / \Gamma(\decay{\Dzb(\tau)}{\bar{f}})$} (for \mbox{$\Gamma(\decay{\Dzb(\tau)}{f}) / \Gamma(\decay{\Dz(\tau)}{f})$}), where the same approximations as above are employed.

The expansion up to second order in the mixing parameters of the time-dependent \CP asymmetry defined in \cref{eq:acp_CS} for \Dz decays into the CS final state $f$ is
\begin{equation}
    \label{eq:acp_CS_quadratic}
    A_{\CP}(f,\tau) \approx \Acpdec{f} + \DY{f}\tau
                 + \frac{1}{2}\Big[x_{12}y_{12}(\sin\phi^M_f - \sin\phi^\Gamma_f)
                 - (x_{12}^2 + y^2_{12})\Acpdec{f}\Big]\tau^2,
\end{equation}
where terms multiplying \Acpdec{f} have been expanded to first order in the \CP-violation parameters \Acpdec{f}, $\sin\phi^M_f$ and $\sin\phi^\Gamma_f$, and terms of order higher than two in $\sin\phi^M_f$ and $\sin\phi^\Gamma_f$ have been neglected.
Since both mixing parameters are of the order of \mbox{$5 \times 10^{-3}$}~\cite{HFLAV18,LHCb-PAPER-2021-033}, \Acpdec{f} is of the order of $1 \times 10^{-3}$~\cite{LHCb-PAPER-2014-013,LHCb-PAPER-2019-006} and the effective maximum value of $\tau$ tested at experiments is of the order of 6~\cite{LHCb-PAPER-2020-045}, the term quadratic in the mixing parameters might become relevant only at large decay times and at the precision targeted for \DY{f} by the \lhcb Upgrade~II, \mbox{$1 \times 10^{-5}$}~\cite{Bediaga:2018lhg}, and if the magnitude of $\phi^M_2$ or $\phi^\Gamma_2$ is of the same order of the current experimental limits, $\mathcal{O}(0.05\rad)$~\cite{LHCb-PAPER-2021-033}.
The latter condition is unlikely in the SM~\cite{Kagan:2020vri,kagan:charm2021}.

\bibliographystyle{JHEP-tpajero}
\bibliography{main,standard,LHCb-PAPER}

\end{document}

%% file: lhcb-symbols-def.tex
\usepackage{xspace} 
\usepackage{upgreek}

\def\lhcb   {\mbox{LHCb}\xspace}

\def\belle  {\mbox{Belle}\xspace}
\def\belletwo {\mbox{Belle~II}\xspace}

\def\MagUp {\mbox{\em Mag\kern -0.05em Up}\xspace}

\ifthenelse{\boolean{uprightparticles}}%
{

 \def\Pmu         {\ensuremath{\upmu}\xspace}

 \def\Ppi         {\ensuremath{\uppi}\xspace}

 \def\Pphi        {\ensuremath{\upphi}\xspace}

 \def\PDelta      {\ensuremath{\Delta}\xspace}                 
 \def\PXi         {\ensuremath{\Xi}\xspace}                 
 \def\PLambda     {\ensuremath{\Lambda}\xspace}                 
 \def\PSigma      {\ensuremath{\Sigma}\xspace}                 
 \def\POmega      {\ensuremath{\Omega}\xspace}                 
 \def\PUpsilon    {\ensuremath{\Upsilon}\xspace}

 \def\PB      {\ensuremath{\mathrm{B}}\xspace}                 
                  
 \def\PD      {\ensuremath{\mathrm{D}}\xspace}

 \def\PK      {\ensuremath{\mathrm{K}}\xspace}

 \def\PP      {\ensuremath{\mathrm{P}}\xspace}

 \def\Ph      {\ensuremath{\mathrm{h}}\xspace}                 
 \def\Pi      {\ensuremath{\mathrm{i}}\xspace}

 \def\Ps      {\ensuremath{\mathrm{s}}\xspace}

 \def\thebaroffset{0.0em}
}
{

 \def\Pmu         {\ensuremath{\mu}\xspace}

 \def\Ppi         {\ensuremath{\pi}\xspace}

 \def\Pphi        {\ensuremath{\phi}\xspace}

 \mathchardef\PDelta="7101
 \mathchardef\PXi="7104
 \mathchardef\PLambda="7103
 \mathchardef\PSigma="7106
 \mathchardef\POmega="710A
 \mathchardef\PUpsilon="7107
                  
 \def\PB      {\ensuremath{B}\xspace}                 
                  
 \def\PD      {\ensuremath{D}\xspace}

 \def\PK      {\ensuremath{K}\xspace}

 \def\PP      {\ensuremath{P}\xspace}

 \def\Ph      {\ensuremath{h}\xspace}                 
 \def\Pi      {\ensuremath{i}\xspace}

 \def\Ps      {\ensuremath{s}\xspace}

 \def\thebaroffset{0.18em}
}
\newcommand{\offsetoverline}[2][\thebaroffset]{\kern #1\overline{\kern -#1 #2}}%

\makeatletter
\ifcase \@ptsize \relax% 10pt
  \newcommand{\miniscule}{\@setfontsize\miniscule{4}{5}}% \tiny: 5/6
\or% 11pt
  \newcommand{\miniscule}{\@setfontsize\miniscule{5}{6}}% \tiny: 6/7
\or% 12pt
  \newcommand{\miniscule}{\@setfontsize\miniscule{5}{6}}% \tiny: 6/7
\fi
\makeatother

\DeclareRobustCommand{\optbar}[1]{\shortstack{{\miniscule (\rule[.5ex]{1.25em}{.18mm})}
  \\ [-.7ex] $#1$}}

\def\mun        {{\ensuremath{\Pmu^-}}\xspace} % muon negative (\mum is taken)

\def\squark    {{\ensuremath{\Ps}}\xspace}

\def\hadron {{\ensuremath{\Ph}}\xspace}
\def\pion   {{\ensuremath{\Ppi}}\xspace}

\def\pip    {{\ensuremath{\pion^+}}\xspace}
\def\pim    {{\ensuremath{\pion^-}}\xspace}

\def\pimp   {{\ensuremath{\pion^\mp}}\xspace}

\def\kaon    {{\ensuremath{\PK}}\xspace}
\def\KorKbar {\kern \thebaroffset\optbar{\kern -\thebaroffset \PK}{}\xspace}

\def\Kp      {{\ensuremath{\kaon^+}}\xspace}
\def\Km      {{\ensuremath{\kaon^-}}\xspace}
\def\Kpm     {{\ensuremath{\kaon^\pm}}\xspace}

\def\KS      {{\ensuremath{\kaon^0_{\mathrm{S}}}}\xspace}

\newcommand{\phiz}{\ensuremath{\Pphi}\xspace}

\def\Dbar    {{\ensuremath{\offsetoverline{\PD}}}\xspace}
\def\D       {{\ensuremath{\PD}}\xspace}

\def\DorDbar {\kern \thebaroffset\optbar{\kern -\thebaroffset \PD}\xspace}
\def\Dz      {{\ensuremath{\D^0}}\xspace}
\def\Dzb     {{\ensuremath{\Dbar{}^0}}\xspace}

\def\theDstarp{{\ensuremath{\D^{*}(2010)^{+}}}\xspace}

\def\B       {{\ensuremath{\PB}}\xspace}
\def\Bbar    {{\ensuremath{\offsetoverline{\PB}}}\xspace}
\def\Bb      {{\ensuremath{\Bbar}}\xspace}
\def\BorBbar {\kern \thebaroffset\optbar{\kern -\thebaroffset \PB}\xspace}

\def\Bd      {{\ensuremath{\B^0}}\xspace}

\def\BdorBdbar {\kern \thebaroffset\optbar{\kern -\thebaroffset \Bd}\xspace}

\def\Bs      {{\ensuremath{\B^0_\squark}}\xspace}

\def\BsorBsbar {\kern \thebaroffset\optbar{\kern -\thebaroffset \Bs}\xspace}

\def\Y#1S{\ensuremath{\PUpsilon{(#1S)}}\xspace}

\def\LorLbar     {\kern \thebaroffset\optbar{\kern -\thebaroffset \PLambda}\xspace}

\newcommand{\decay}[2]{\ensuremath{#1\!\to #2}\xspace} 

\def\to                 {\ensuremath{\rightarrow}\xspace}

\def\CP                {{\ensuremath{C\!P}}\xspace}
\def\CPT               {{\ensuremath{C\!PT}}\xspace}

\def\AT#1     {\ensuremath{A_{\mathrm{T}}^{#1}}\xspace}           % 2

\def\C#1      {\ensuremath{\mathcal{C}_{#1}}\xspace}                       % 9
\def\Cp#1     {\ensuremath{\mathcal{C}_{#1}^{'}}\xspace}                    % 7
\def\Ceff#1   {\ensuremath{\mathcal{C}_{#1}^{\mathrm{(eff)}}}\xspace}        % 9  
\def\Cpeff#1  {\ensuremath{\mathcal{C}_{#1}^{'\mathrm{(eff)}}}\xspace}       % 7
\def\Ope#1    {\ensuremath{\mathcal{O}_{#1}}\xspace}                       % 2
\def\Opep#1   {\ensuremath{\mathcal{O}_{#1}^{'}}\xspace}                    % 7

\newcommand{\bra}[1]{\ensuremath{\langle #1|}}             % {a}
\newcommand{\ket}[1]{\ensuremath{|#1\rangle}}              % {b}
\newcommand{\aunit}[1]{\ensuremath{\text{\,#1}}}       
\newcommand{\tev}{\aunit{Te\kern -0.1em V}\xspace}
\newcommand{\gev}{\aunit{Ge\kern -0.1em V}\xspace}
\newcommand{\mev}{\aunit{Me\kern -0.1em V}\xspace}
\newcommand{\kev}{\aunit{ke\kern -0.1em V}\xspace}
\newcommand{\ev}{\aunit{e\kern -0.1em V}\xspace}
\newcommand{\mevc}{\ensuremath{\aunit{Me\kern -0.1em V\!/}c}\xspace}
\newcommand{\gevc}{\ensuremath{\aunit{Ge\kern -0.1em V\!/}c}\xspace}
\newcommand{\mevcc}{\ensuremath{\aunit{Me\kern -0.1em V\!/}c^2}\xspace}
\newcommand{\gevcc}{\ensuremath{\aunit{Ge\kern -0.1em V\!/}c^2}\xspace}
\def\gsim{{~\raise.15em\hbox{$>$}\kern-.85em
          \lower.35em\hbox{$\sim$}~}\xspace}
\def\lsim{{~\raise.15em\hbox{$<$}\kern-.85em
          \lower.35em\hbox{$\sim$}~}\xspace}

\def\mrad{\aunit{mrad}}
\def\rad{\aunit{rad}}

\def\tell1  {TELL1\xspace}
\def\ukl1   {UKL1\xspace}

%% file: my-def.tex
\def\af {{\ensuremath{A_f}}\xspace}
\def\afb {{\ensuremath{A_{\bar{f}}}}\xspace}
\def\abf {{\ensuremath{\bar{A}_f}}\xspace}
\def\abfb {{\ensuremath{\bar{A}_{\bar{f}}}}\xspace}

\def\DzRS {\decay{\Dz}{\Km\pip}}

\def\DzKK {\decay{\Dz}{\Kp\Km}}
\def\DzPP {\decay{\Dz}{\pip\pim}}

\def\hp     {{\ensuremath{\hadron^+}}\xspace}
\def\hm     {{\ensuremath{\hadron^-}}\xspace}

\def\KK {{\ensuremath{\Kp\Km  }}\xspace}
\def\PP {{\ensuremath{\pip\pim}}\xspace}
\def\RS {{\ensuremath{\Km\pip }}\xspace}
\def\WS {{\ensuremath{\Kp\pim }}\xspace}

\newcommand{\DY}[1]{{\ensuremath{\Delta Y_{#1}}}\xspace}
\newcommand{\agamma}[1]{{\ensuremath{A_{\Gamma}^{#1}}}\xspace}
\newcommand{\ycp}[1]{{\ensuremath{y_{\CP}^{#1}}}\xspace}
\newcommand{\Acpdec}[1]{{\ensuremath{a^{d}_{#1}}}\xspace}

\newcommand{\etacp}[1]{{\ensuremath{\eta^{\CP}_{#1}}}\xspace}

%% file: main.bbl
\providecommand{\href}[2]{#2}\begingroup\raggedright\begin{thebibliography}{10}

\bibitem{Grossman:2006jg}
Y.~Grossman, A.L.~Kagan and Y.~Nir, \emph{{New physics and CP violation in
  singly Cabibbo suppressed D decays}},
  \href{https://doi.org/10.1103/PhysRevD.75.036008}{\emph{Phys. Rev. D}
  {\bfseries 75} (2007) 036008}
  [\href{https://arxiv.org/abs/hep-ph/0609178}{{\ttfamily hep-ph/0609178}}].

\bibitem{Aubert:2007wf}
{\scshape BaBar} collaboration, \emph{{Evidence for $D^{\,0} -
  \overline{D}^{\,0}$ mixing}},
  \href{https://doi.org/10.1103/PhysRevLett.98.211802}{\emph{Phys. Rev. Lett.}
  {\bfseries 98} (2007) 211802}
  [\href{https://arxiv.org/abs/hep-ex/0703020}{{\ttfamily hep-ex/0703020}}].

\bibitem{delAmoSanchez:2010xz}
{\scshape BaBar} collaboration, \emph{{Measurement of $\Dz-\Dzb$ mixing
  parameters using \decay{\Dz}{\KS\pip\pim} and \decay{\Dz}{\KS\Kp\Km}
  decays}}, \href{https://doi.org/10.1103/PhysRevLett.105.081803}{\emph{Phys.
  Rev. Lett.} {\bfseries 105} (2010) 081803}
  [\href{https://arxiv.org/abs/1004.5053}{{\ttfamily arXiv:1004.5053}}].

\bibitem{Lees:2012qh}
{\scshape BaBar} collaboration, \emph{{Measurement of \Dz--\Dzb mixing and CP
  violation in two-body \Dz decays}},
  \href{https://doi.org/10.1103/PhysRevD.87.012004}{\emph{Phys. Rev. D}
  {\bfseries 87} (2013) 012004}
  [\href{https://arxiv.org/abs/1209.3896}{{\ttfamily arXiv:1209.3896}}].

\bibitem{Aaltonen:2013pja}
{\scshape CDF} collaboration, \emph{{Observation of $D^0$--$\bar D^0$ mixing
  using the CDF II detector}},
  \href{https://doi.org/10.1103/PhysRevLett.111.231802}{\emph{Phys. Rev. Lett.}
  {\bfseries 111} (2013) 231802}
  [\href{https://arxiv.org/abs/1309.4078}{{\ttfamily arXiv:1309.4078}}].

\bibitem{Ko:2014qvu}
{\scshape Belle} collaboration, \emph{{Observation of $D^0-\bar{D}^0$ mixing in
  $e^+e^-$ collisions}},
  \href{https://doi.org/10.1103/PhysRevLett.112.111801}{\emph{Phys. Rev. Lett.}
  {\bfseries 112} (2014) 111801} [Erratum
  \href{https://doi.org/10.1103/PhysRevLett.112.139903}{ ibid\   \textbf{112}
  (2014) 139903}] [\href{https://arxiv.org/abs/1401.3402}{{\ttfamily
  arXiv:1401.3402}}].

\bibitem{Peng:2014oda}
{\scshape Belle} collaboration, \emph{{Measurement of $D^0-\bar{D}^0$ mixing
  and search for indirect CP violation using $D^0\to K_S^0\pi^+\pi^-$ decays}},
  \href{https://doi.org/10.1103/PhysRevD.89.091103}{\emph{Phys. Rev. D}
  {\bfseries 89} (2014) 091103}
  [\href{https://arxiv.org/abs/1404.2412}{{\ttfamily arXiv:1404.2412}}].

\bibitem{Staric:2015sta}
{\scshape Belle} collaboration, \emph{{Measurement of \Dz--\Dzb mixing and
  search for CP violation in $D^0 \to K^+ K^-, \pi^+ \pi^-$ decays with the
  full Belle data set}},
  \href{https://doi.org/10.1016/j.physletb.2015.12.025}{\emph{Phys. Lett. B}
  {\bfseries 753} (2016) 412}
  [\href{https://arxiv.org/abs/1509.08266}{{\ttfamily arXiv:1509.08266}}].

\bibitem{LHCb-PAPER-2015-057}
LHCb collaboration, \emph{{First observation of $\Dz-\Dzb$ oscillations in
  \mbox{\decay{\Dz}{\Kp\pip\pim\pim}} decays and a measurement of the
  associated coherence parameters}},
  \href{https://doi.org/10.1103/PhysRevLett.116.241801}{\emph{Phys. Rev. Lett.}
  {\bfseries 116} (2016) 241801}
  [\href{https://arxiv.org/abs/1602.07224}{{\ttfamily arXiv:1602.07224}}].

\bibitem{BaBar:2016kvp}
{\scshape BaBar} collaboration, \emph{{Measurement of the neutral $D$ meson
  mixing parameters in a time-dependent amplitude analysis of the
  $D^0\to\pi^+\pi^-\pi^0$ decay}},
  \href{https://doi.org/10.1103/PhysRevD.93.112014}{\emph{Phys. Rev. D}
  {\bfseries 93} (2016) 112014}
  [\href{https://arxiv.org/abs/1604.00857}{{\ttfamily arXiv:1604.00857}}].

\bibitem{LHCb-PAPER-2016-033}
LHCb collaboration, \emph{{Measurements of charm mixing and \CP violation using
  \mbox{\decay{\Dz}{\Kpm\pimp}} decays}},
  \href{https://doi.org/10.1103/PhysRevD.95.052004}{\emph{Phys. Rev. D}
  {\bfseries 95} (2017) 052004} [Erratum
  \href{https://doi.org/10.1103/PhysRevD.96.099907}{ ibid\   \textbf{96} (2017)
  099907}] [\href{https://arxiv.org/abs/1611.06143}{{\ttfamily
  arXiv:1611.06143}}].

\bibitem{LHCb-PAPER-2017-046}
LHCb collaboration, \emph{{Updated determination of \Dz--\Dzb mixing and \CP
  violation parameters with \mbox{\decay{\Dz}{\Kp\pim}} decays}},
  \href{https://doi.org/10.1103/PhysRevD.97.031101}{\emph{Phys. Rev. D}
  {\bfseries 97} (2018) 031101}
  [\href{https://arxiv.org/abs/1712.03220}{{\ttfamily arXiv:1712.03220}}].

\bibitem{LHCb-PAPER-2018-038}
LHCb collaboration, \emph{{Measurement of the charm-mixing parameter $y_\CP$}},
  \href{https://doi.org/10.1103/PhysRevLett.122.011802}{\emph{Phys. Rev. Lett.}
  {\bfseries 122} (2019) 011802}
  [\href{https://arxiv.org/abs/1810.06874}{{\ttfamily arXiv:1810.06874}}].

\bibitem{LHCb-PAPER-2019-001}
LHCb collaboration, \emph{{Measurement of the mass difference between neutral
  charm-meson eigenstates}},
  \href{https://doi.org/10.1103/PhysRevLett.122.231802}{\emph{Phys. Rev. Lett.}
  {\bfseries 122} (2019) 231802}
  [\href{https://arxiv.org/abs/1903.03074}{{\ttfamily arXiv:1903.03074}}].

\bibitem{LHCb-PAPER-2021-009}
LHCb collaboration, \emph{{Observation of the mass difference between neutral
  charm-meson eigenstates}},
  \href{https://doi.org/10.1103/PhysRevLett.127.111801}{\emph{Phys. Rev. Lett.}
  {\bfseries 127} (2021) 111801}
  [\href{https://arxiv.org/abs/2106.03744}{{\ttfamily arXiv:2106.03744}}].

\bibitem{LHCb-PAPER-2021-033}
LHCb collaboration, \emph{{Simultaneous determination of CKM~angle~$\gamma$ and
  charm mixing parameters}},
  \href{https://doi.org/10.1007/JHEP12(2021)141}{\emph{JHEP} {\bfseries 12}
  (2021) 141} [\href{https://arxiv.org/abs/2110.02350}{{\ttfamily
  arXiv:2110.02350}}].

\bibitem{LHCb-PAPER-2019-006}
LHCb collaboration, \emph{{Observation of \CP violation in charm decays}},
  \href{https://doi.org/10.1103/PhysRevLett.122.211803}{\emph{Phys. Rev. Lett.}
  {\bfseries 122} (2019) 211803}
  [\href{https://arxiv.org/abs/1903.08726}{{\ttfamily arXiv:1903.08726}}].

\bibitem{Aaltonen:2014efa}
{\scshape CDF} collaboration, \emph{{Measurement of indirect CP-violating
  asymmetries in $D^0\to K^+K^-$ and $D^0\to \pi^+\pi^-$ decays at CDF}},
  \href{https://doi.org/10.1103/PhysRevD.90.111103}{\emph{Phys. Rev. D}
  {\bfseries 90} (2014) 111103}
  [\href{https://arxiv.org/abs/1410.5435}{{\ttfamily arXiv:1410.5435}}].

\bibitem{LHCb-PAPER-2014-069}
LHCb collaboration, \emph{{Measurement of indirect \CP asymmetries in
  \mbox{\decay{\Dz}{\Km\Kp}} and \mbox{\decay{\Dz}{\pim\pip}} decays using
  semileptonic \B decays}},
  \href{https://doi.org/10.1007/JHEP04(2015)043}{\emph{JHEP} {\bfseries 04}
  (2015) 043} [\href{https://arxiv.org/abs/1501.06777}{{\ttfamily
  arXiv:1501.06777}}].

\bibitem{LHCb-PAPER-2016-063}
LHCb collaboration, \emph{{Measurement of the \CP violation parameter
  $A_\Gamma$ in \mbox{\decay{\Dz}{\Kp\Km}} and \mbox{\decay{\Dz}{\pip\pim}}
  decays}}, \href{https://doi.org/10.1103/PhysRevLett.118.261803}{\emph{Phys.
  Rev. Lett.} {\bfseries 118} (2017) 261803}
  [\href{https://arxiv.org/abs/1702.06490}{{\ttfamily arXiv:1702.06490}}].

\bibitem{LHCb-PAPER-2019-032}
LHCb collaboration, \emph{{Updated measurement of decay-time-dependent \CP
  asymmetries in \mbox{\decay{\Dz}{\Kp\Km}} and \mbox{\decay{\Dz}{\pip\pim}}
  decays}}, \href{https://doi.org/10.1103/PhysRevD.101.012005}{\emph{Phys. Rev.
  D} {\bfseries 101} (2020) 012005}
  [\href{https://arxiv.org/abs/1911.01114}{{\ttfamily arXiv:1911.01114}}].

\bibitem{LHCb-PAPER-2020-045}
LHCb collaboration, \emph{{Search for time-dependent \CP violation in $\Dz \to
  \Kp \Km$ and $\Dz \to \pip \pim$ decays}},
  \href{https://doi.org/10.1103/PhysRevD.104.072010}{\emph{Phys. Rev. D}
  {\bfseries 104} (2021) 072010}
  [\href{https://arxiv.org/abs/2105.09889}{{\ttfamily arXiv:2105.09889}}].

\bibitem{Bediaga:2018lhg}
LHCb collaboration, \emph{Physics case for an {LHCb} \mbox{Upgrade II} --
  {O}pportunities in flavour physics, and beyond, in the {HL-LHC} era},  2018,
  pp. 51--57 [\href{https://arxiv.org/abs/1808.08865}{{\ttfamily
  arXiv:1808.08865}}].

\bibitem{Kou:2018nap}
{\scshape Belle II} collaboration, \emph{{The Belle II physics book}},
  \href{https://doi.org/10.1093/ptep/ptz106}{\emph{PTEP} {\bfseries 2019}
  123C01} [Erratum \href{https://doi.org/10.1093/ptep/ptaa008}{ ibid\
  \textbf{2020} (2020) 029201}]
  [\href{https://arxiv.org/abs/1808.10567}{{\ttfamily arXiv:1808.10567}}], pp.
  379--388.

\bibitem{Bigi:2011re}
I.I.~Bigi, A.~Paul and S.~Recksiegel, \emph{{Conclusions from CDF results on
  $C\!P$ violation in $D^0 \to \pi^+\pi^-, K^+K^-$ and future tasks}},
  \href{https://doi.org/10.1007/JHEP06(2011)089}{\emph{JHEP} {\bfseries 06}
  (2011) 089} [\href{https://arxiv.org/abs/1103.5785}{{\ttfamily
  arXiv:1103.5785}}].

\bibitem{Bobrowski:2010xg}
M.~Bobrowski, A.~Lenz, J.~Riedl and J.~Rohrwild, \emph{{How large can the SM
  contribution to $C\!P$ violation in $D^0-\bar D^0$ mixing be?}},
  \href{https://doi.org/10.1007/JHEP03(2010)009}{\emph{JHEP} {\bfseries 03}
  (2010) 009} [\href{https://arxiv.org/abs/1002.4794}{{\ttfamily
  arXiv:1002.4794}}].

\bibitem{Kagan:2020vri}
A.L.~Kagan and L.~Silvestrini, \emph{{Dispersive and absorptive $C\!P$
  violation in $D^0- \overline{D^0}$ mixing}},
  \href{https://doi.org/10.1103/PhysRevD.103.053008}{\emph{Phys. Rev. D}
  {\bfseries 103} (2021) 053008}
  [\href{https://arxiv.org/abs/2001.07207}{{\ttfamily arXiv:2001.07207}}].

\bibitem{Li:2020xrz}
H.-N.~Li, H.~Umeeda, F.~Xu and F.-S.~Yu, \emph{{$D$ meson mixing as an inverse
  problem}}, \href{https://doi.org/10.1016/j.physletb.2020.135802}{\emph{Phys.
  Lett. B} {\bfseries 810} (2020) 135802}
  [\href{https://arxiv.org/abs/2001.04079}{{\ttfamily arXiv:2001.04079}}].

\bibitem{kagan:charm2021}
A.L.~Kagan, \emph{{Describing charm time-dependent $C\!P$ violation in the
  precision era}},  {talk at the \textit{10$^\text{th}$ International workshop
  on charm physics}, Mexico City, Mexico}, 31 May--4 June 2021, {online at
  \href{https://indico.nucleares.unam.mx/event/1488/session/17/contribution/66}{https://indico.nucleares.unam.mx/event/1488/session/17/contribution/66}}.

\bibitem{Xing:1996pn}
Z.-Z.~Xing, \emph{{\Dz-\Dzb mixing and \CP violation in neutral \D-meson
  decays}}, \href{https://doi.org/10.1103/PhysRevD.55.196}{\emph{Phys. Rev. D}
  {\bfseries 55} (1997) 196}
  [\href{https://arxiv.org/abs/hep-ph/9606422}{{\ttfamily hep-ph/9606422}}].

\bibitem{Grossman:2009mn}
Y.~Grossman, Y.~Nir and G.~Perez, \emph{{Testing new indirect $C\!P$
  violation}},
  \href{https://doi.org/10.1103/PhysRevLett.103.071602}{\emph{Phys. Rev. Lett.}
  {\bfseries 103} (2009) 071602}
  [\href{https://arxiv.org/abs/0904.0305}{{\ttfamily arXiv:0904.0305}}].

\bibitem{Kagan:2009gb}
A.L.~Kagan and M.D.~Sokoloff, \emph{{Indirect CP violation and implications for
  \Dz--\Dzb and \Bs--\Bsb mixing}},
  \href{https://doi.org/10.1103/PhysRevD.80.076008}{\emph{Phys. Rev. D}
  {\bfseries 80} (2009) 076008}
  [\href{https://arxiv.org/abs/0907.3917}{{\ttfamily arXiv:0907.3917}}].

\bibitem{HFLAV18}
{\scshape Heavy Flavor Averaging Group}, \emph{{Averages of $b$-hadron,
  $c$-hadron, and $\tau$-lepton properties as of 2018}},
  \href{https://doi.org/10.1140/epjc/s10052-020-8156-7}{\emph{Eur. Phys. J. C}
  {\bfseries 81} (2021) 226}
  [\href{https://arxiv.org/abs/1909.12524}{{\ttfamily arXiv:1909.12524}}],
  {updated results and plots available online at
  \href{https://hflav.web.cern.ch}{https://hflav.web.cern.ch}}.

\bibitem{pajero:2021}
T.~Pajero, \emph{{Search for time-dependent \CP violation in \DzKK and \DzPP
  decays}}, Ph.D. thesis, Scuola Normale Superiore, Pisa, Italy, 2021
  [\href{https://cds.cern.ch/record/2747731/}{CERN-THESIS-2020-231}].

\bibitem{LHCb-PAPER-2013-054}
LHCb collaboration, \emph{{Measurements of indirect \CP asymmetries in
  \mbox{\decay{\Dz}{\Km\Kp}} and \mbox{\decay{\Dz}{\pim\pip}} decays}},
  \href{https://doi.org/10.1103/PhysRevLett.112.041801}{\emph{Phys. Rev. Lett.}
  {\bfseries 112} (2014) 041801}
  [\href{https://arxiv.org/abs/1310.7201}{{\ttfamily arXiv:1310.7201}}].

\bibitem{pajero:code}
{T. Pajero}, \emph{{Fitter of charm mixing and \CP violation parameters}},
  {October} {2021}, online at
  \href{https://github.com/tpajero/charm-fitter}{https://github.com/tpajero/charm-fitter}.

\bibitem{PDG2020}
{\scshape Particle Data Group}, \emph{Review of particle physics},
  \href{https://doi.org/10.1093/ptep/ptaa104}{\emph{Prog. Theor. Exp. Phys.}
  {\bfseries 2020} (2020) 083C01}, {updated results available online at
  \href{https://pdg.lbl.gov/}{https://pdg.lbl.gov/}}.

\bibitem{LHCb-CONF-2019-001}
LHCb collaboration, \emph{{Search for time-dependent \CP violation in
  \decay{\Dz}{\Kp\Km} and \decay{\Dz}{\pip\pim} decays}},  May 2019,
  [\href{https://lhcbproject.web.cern.ch/lhcbproject/Publications/LHCbProjectPublic/LHCb-CONF-2019-001.html}{LHCb-CONF-2019-001}].

\bibitem{LHCb-PAPER-2014-013}
LHCb collaboration, \emph{{Measurement of \CP asymmetry in
  \mbox{\decay{\Dz}{\Km\Kp}} and \mbox{\decay{\Dz}{\pim\pip}} decays}},
  \href{https://doi.org/10.1007/JHEP07(2014)041}{\emph{JHEP} {\bfseries 07}
  (2014) 041} [\href{https://arxiv.org/abs/1405.2797}{{\ttfamily
  arXiv:1405.2797}}].

\bibitem{LHCb-PAPER-2012-041}
LHCb collaboration, \emph{{Prompt charm production in \proton\proton collisions
  at \mbox{$\sqs=$7\tev}}},
  \href{https://doi.org/10.1016/j.nuclphysb.2013.02.010}{\emph{Nucl. Phys. B}
  {\bfseries 871} (2013) 1} [\href{https://arxiv.org/abs/1302.2864}{{\ttfamily
  arXiv:1302.2864}}].

\bibitem{LHCb-PAPER-2015-041}
LHCb collaboration, \emph{{Measurements of prompt charm production
  cross-sections in \proton\proton collisions at $\sqs = $13\tev}},
  \href{https://doi.org/10.1007/JHEP03(2016)159}{\emph{JHEP} {\bfseries 03}
  (2016) 159} [Erratum \href{https://doi.org/10.1007/JHEP09(2016)013}{ ibid\
  \textbf{09} (2016) 013}] [Erratum
  \href{https://doi.org/10.1007/JHEP05(2017)074}{ ibid\   \textbf{05} (2017)
  074}] [\href{https://arxiv.org/abs/1510.01707}{{\ttfamily
  arXiv:1510.01707}}].

\bibitem{LHCb-CONF-2016-005}
LHCb collaboration, \emph{{LHCb dimuon and charm mass distributions}},  July
  2016,
  [\href{https://lhcbproject.web.cern.ch/lhcbproject/Publications/LHCbProjectPublic/LHCb-CONF-2016-005.html}{LHCb-CONF-2016-005}].

\bibitem{E791:1999bzz}
{\scshape E791} collaboration, \emph{{Measurements of lifetimes and a limit on
  the lifetime difference in the neutral D meson system}},
  \href{https://doi.org/10.1103/PhysRevLett.83.32}{\emph{Phys. Rev. Lett.}
  {\bfseries 83} (1999) 32}
  [\href{https://arxiv.org/abs/hep-ex/9903012}{{\ttfamily hep-ex/9903012}}].

\end{thebibliography}\endgroup
